\documentclass[manuscript]{aastex62}
\usepackage[encapsulated]{CJK}
\usepackage{mathrsfs}
\usepackage{subfigure}
\usepackage{amsmath}
\usepackage{longtable}
\usepackage{booktabs}
\usepackage{makecell}
\usepackage{threeparttable}
\usepackage{hyperref} 
\usepackage{overpic}
\usepackage{diagbox}
\usepackage{array}
\usepackage[T1]{fontenc}
\def\etal {et al.~}

\newbox\grsign \setbox\grsign=\hbox{$>$} \newdimen\grdimen \grdimen=\ht\grsign
\newbox\laxbox \newbox\gaxbox
\setbox\gaxbox=\hbox{\raise.5ex\hbox{$>$}\llap
     {\lower.5ex\hbox{$\sim$}}}\ht1=\grdimen\dp1=0pt
\setbox\laxbox=\hbox{\raise.5ex\hbox{$<$}\llap
     {\lower.5ex\hbox{$\sim$}}}\ht2=\grdimen\dp2=0pt

\shorttitle{HI Wind}
\shortauthors{Li \etal}

\definecolor{malachite}{rgb}{0.34, 0.7, 0.22}

\begin{document}
\begin{CJK*}{UTF8}{gbsn}

\title{Neutral Stellar Winds Toward the High-Mass Star-Forming Region G176.51+00.20}

\correspondingauthor{Yingjie Li}
\email{liyj@pmo.ac.cn, xuye@pmo.ac.cn}

\author{Yingjie Li}\affiliation{Purple Mountain Observatory, Chinese Academy of Sciences, Nanjing 210008, China}

\author{Ye Xu}
\affiliation{Purple Mountain Observatory, Chinese Academy of Sciences, Nanjing 210008, China}

\author{Jin-Long Xu}
\affiliation{National Astronomical Observatories, Chinese Academy of Sciences, Beijing 100101, China}
\affiliation{CAS Key Laboratory of FAST, National Astronomical Observatories, Chinese Academy of Sciences, Beijing 100101, China}

\author{Dejian Liu}
\affiliation{Purple Mountain Observatory, Chinese Academy of Sciences, Nanjing 210008, China}
\affiliation{University of Science and Technology of China, Hefei, Anhui 230026, China}

\author{Jingjing Li}
\affiliation{Purple Mountain Observatory, Chinese Academy of Sciences, Nanjing 210008, China}

\author{Zehao Lin}
\affiliation{Purple Mountain Observatory, Chinese Academy of Sciences, Nanjing 210008, China}
\affiliation{University of Science and Technology of China, Hefei, Anhui 230026, China}

\author{Peng Jiang}
\affiliation{National Astronomical Observatories, Chinese Academy of Sciences, Beijing 100101, China}
\affiliation{CAS Key Laboratory of FAST, National Astronomical Observatories, Chinese Academy of Sciences, Beijing 100101, China}

\author{Shuaibo Bian}
\affiliation{Purple Mountain Observatory, Chinese Academy of Sciences, Nanjing 210008, China}
\affiliation{University of Science and Technology of China, Hefei, Anhui 230026, China}

\author{Chaojie Hao}
\affiliation{Purple Mountain Observatory, Chinese Academy of Sciences, Nanjing 210008, China}
\affiliation{University of Science and Technology of China, Hefei, Anhui 230026, China}

\author{Xiuhui Chen}
\affiliation{College of Mathematics and Physics, Hunan University of Arts and Science, Changde, Hunan 415300, China}

\begin{abstract}
We observed the high-mass star-forming region G176.51+00.20 using the Five-hundred-meter Aperture Spherical radio Telescope (FAST) with the 19-beam tracking observational mode. This is a pilot work of searching for neutral stellar winds traced by atomic hydrogen (i.e., H~\textsc{i} winds) using the high sensitivity H~\textsc{i} line toward high-mass star-forming regions where bipolar molecular outflows have been detected with high sensitivity by Liu et al. H~\textsc{i} wind was detected in this work only in Beam 1. We find here that, similar to low-mass star formation, no matter how large the inclination is, the H~\textsc{i} wind is likely sufficiently strong to drive a molecular outflow. We also find that the abundance of H~\textsc{i} in the H~\textsc{i} wind is consistent with that of the H~\textsc{i} narrow-line self-absorption (HINSA) in the same beam (i.e., Beam 1). This implies that there is probably an internal relationship between H~\textsc{i} winds and HINSA. This result also reinforces the assertion that H~\textsc{i} winds and detected molecular outflows are associated with each other. 
\end{abstract}



\section{Introduction}

One of the important goals of contemporary astrophysics is to understand star formation \citep[e.g.,][]{Tan+2014, Bally2016}. Relative to low-mass star formation, the formation mechanism(s) of high-mass stars remain poorly understood \citep{Shu+1987, McKee-Ostriker2007, Tan+2014}, as too are the driving scenario(s) of massive molecular outflows, an essential phase of early high-mass star formation \citep{Arce+2007}.

Early studies, especially in the 1980s and 1990s, attempted to ascertain the driving source and mechanism of molecular outflows, mostly for low-mass stars. \citet{Snell+1985} and \cite{Strom+1986} found that the
mass-loss rate of ionized gas is too low to drive CO lobes, and suggested that the stellar winds which drive the CO lobes are largely comprised of neutral gas. \cite{Lizano+1988} carefully compared the neutral stellar wind traced by atomic hydrogen (H~\textsc{i} wind hereafter) detected in HH 7--11 and the corresponding CO/HCO$^+$ outflows, and found that the H~\textsc{i} wind was strong enough to drive the molecular outflows. A similar conclusion was drawn by carefully comparing the H~\textsc{i} winds in both HH 7--11 and L1551 with the corresponding molecular outflows \citep[see][]{Giovanardi+1992}. 
In such studies, which were performed with the 305 m radio telescope of the Arecibo Observatory, the root mean square (rms) noise was typically $\sim$ 3.5--8.5 mK @ 3.9 km s$^{-1}$ with beam size of $3'.2$--$3'.5$. A model based on the detected H~\textsc{i} wind in HH 7--11 was developed by \citet{Lizano+1988} to explain the entrainment of ambient molecular gas (i.e., driving the molecular outflow) by the neutral wind. 
These studies suggested that H~\textsc{i} winds are a promising driving source of molecular outflows. However, is such an assertion also true for massive stars?

To answer this question, samples of the molecular outflows have been assembled. For example, $^{12}$CO, $^{13}$CO, HCO$^+$, and CS outflowing gases have been recently detected with high sensitivity (with main beam rms noise of dozens of mK) toward nine high-mass star-forming regions using the 13.7 m millimeter telescope of the Purple Mountain Observatory in Delingha \citep[see more details in][]{Liu+2021}.
Accordingly, an rms noise of $\sim$ 3 mK @ 1.0 km s$^{-1}$ can be achieved with about six hour integrations and a beam size of $\sim 2'.9$ by using the powerful 21 cm H~\textsc{i} line detected by the most sensitive ground-based, single-dish Five-hundred-meter Aperture Spherical radio Telescope \footnote{\url{https://fast.bao.ac.cn/}}\citep[FAST;][]{Nan2006, Nan+2011, Jiang+2020}. The use of FAST will enabled us to conduct broad investigations of stellar winds toward star-forming regions where molecular outflows have been detected.
As such, we have carried out a pilot survey of the H~\textsc{i} wind toward detected molecular outflows with highly sensitive observations. The selected object in this work is one of the nine high-mass star-forming regions in \cite{Liu+2021}, G176.51+00.20.

G176.51+00.20, also known as AFGL 5157 and IRAS 05345+3157, is located 1.8 kpc from Earth \citep{Moffat+1979, Snell+1988}.  
It is a complex region containing multi-generational star formation \citep[e.g.,][]{Chen+2003, Dewangan2019}. The main research object in this work is a high-mass star-forming region at a relatively young phase, where H$_2$O masers and an H~\textsc{ii} region have been detected \citep{Torrelles+1992}. This region is marked as an ``H~\textsc{ii} region'' in figure 1(a) in \citet{Dewangan2019}. 
There is a dense NH$_3$ core in the center of this region, which has become synonymous with this region \citep[e.g.,][]{Torrelles+1992, Chen+2003, Jiang+2013}. By constructing a 3.6 cm map with the Very Large Array, the excitation source of this region was identified as a zero-age main-sequence B3 star \citep{Torrelles+1992b}. The bolometric luminosity of this excitation source was determined as 1.7 $\times$ 10$^3$ $L_{\odot}$ via an 8--1200 $\mu$m SED fit \citep[see][]{Molinari+2008}. The age of this star-forming region was suggested to be $\sim$ 2 $\times$ 10$^{5}$ yr by investigating the near-infrared H$_2$ line emission using the 1.88 m telescope of Okayama Astronomical Observatory, Japan \citep[see][]{Chen+2003}.

The famous bipolar outflow (traced by $^{12}$CO) detected by \citet{Snell+1988} is centered on the dense NH$_3$ core \citep{Molinari+2002, Zhang+2005, Dewangan2019}, suggesting that the excitation source of this bipolar outflow is probably the zero-age main-sequence B3 star \citep[see][]{Torrelles+1992b}. \citet{Liu+2021}  confirmed this bipolar outflow with high-sensitivity $^{12}$CO, $^{13}$CO, C$^{18}$O, HCO$^+$, and CS line emission observations, and broadened the blue lobe of this outflow (traced by $^{12}$CO) by a factor of $\sim$ 2.
This makes the dense NH$_3$ core more suitable to investigate whether its H~\textsc{i} wind is strong enough to drive the molecular outflows.

The remainder of this paper is organized as follows. In Section \ref{sec-data-HI}, we describe the data used in this work. Section \ref{sec-HI-flow} analyses the H~\textsc{i} wind associated with the molecular outflows. In Section \ref{sec-discussion}, discussion of whether the H~\textsc{i} wind is strong enough to drive the  molecular outflows is presented. Finally, Section \ref{sec-summary} gives a summary and the main conclusions of this work.

\section{H~\textsc{i} Observations with FAST} \label{sec-data-HI}

FAST is located in Guizhou Province of southwest China \citep{Nan+2011}. It is equipped with a 19-beam receiver (the frequency range is 1.0--1.5 GHz with a bandwidth of 500 MHz) and has dual linear polarizations \citep[i.e., XX and YY; see][]{Lidi+2018, Jiang+2019, Jiang+2020}. The spectral resolution is $\sim$ 477 Hz, corresponding to a velocity resolution of $\sim$ 0.1 km s$^{-1}$ at 1.4 GHz. The half-power beam width (HPBW) is $\sim$ 2.9$'$ at 1.4 GHz, and the pointing error is $\sim$ 0.2$'$ \citep[see][]{Jiang+2019, Jiang+2020}. 

\begin{figure*}[!htb]
	\centering
    \includegraphics[height=0.45\textwidth]{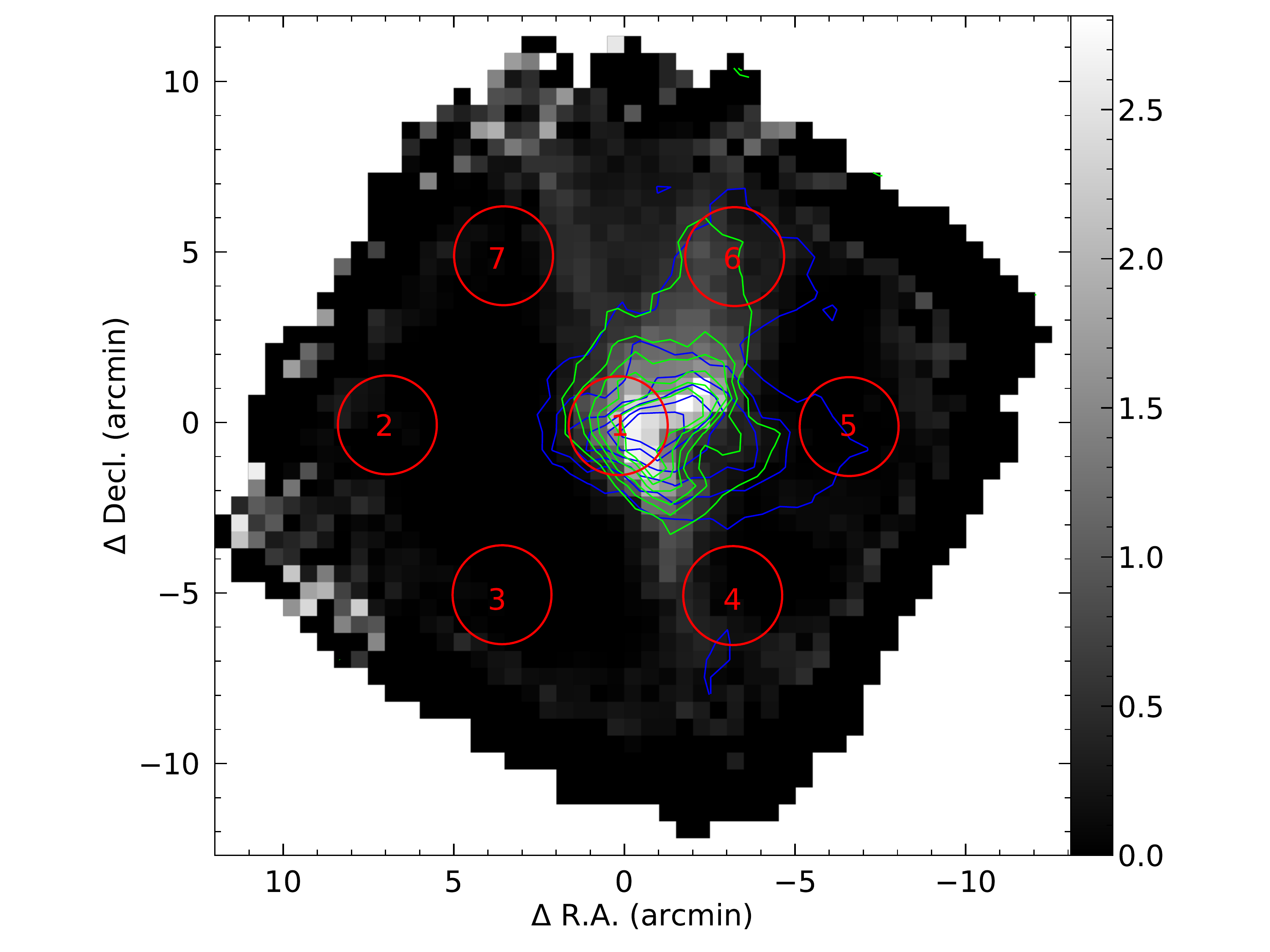}
	\caption{The central seven beams superposed on the integrated intensity map of C$^{18}$O in the range of [$-$30, $-$10] km s$^{-1}$ with the molecular data from \citet{Liu+2021}. The blue and lime contours show the integrated intensity map of $^{12}$CO and $^{13}$CO in the same velocity range as the C$^{18}$O map.}
	\label{figure-positions}
\end{figure*}

\begin{figure*}[!htb]
	\centering
	\subfigure[Beam 1]{\includegraphics[height=0.245\textwidth]{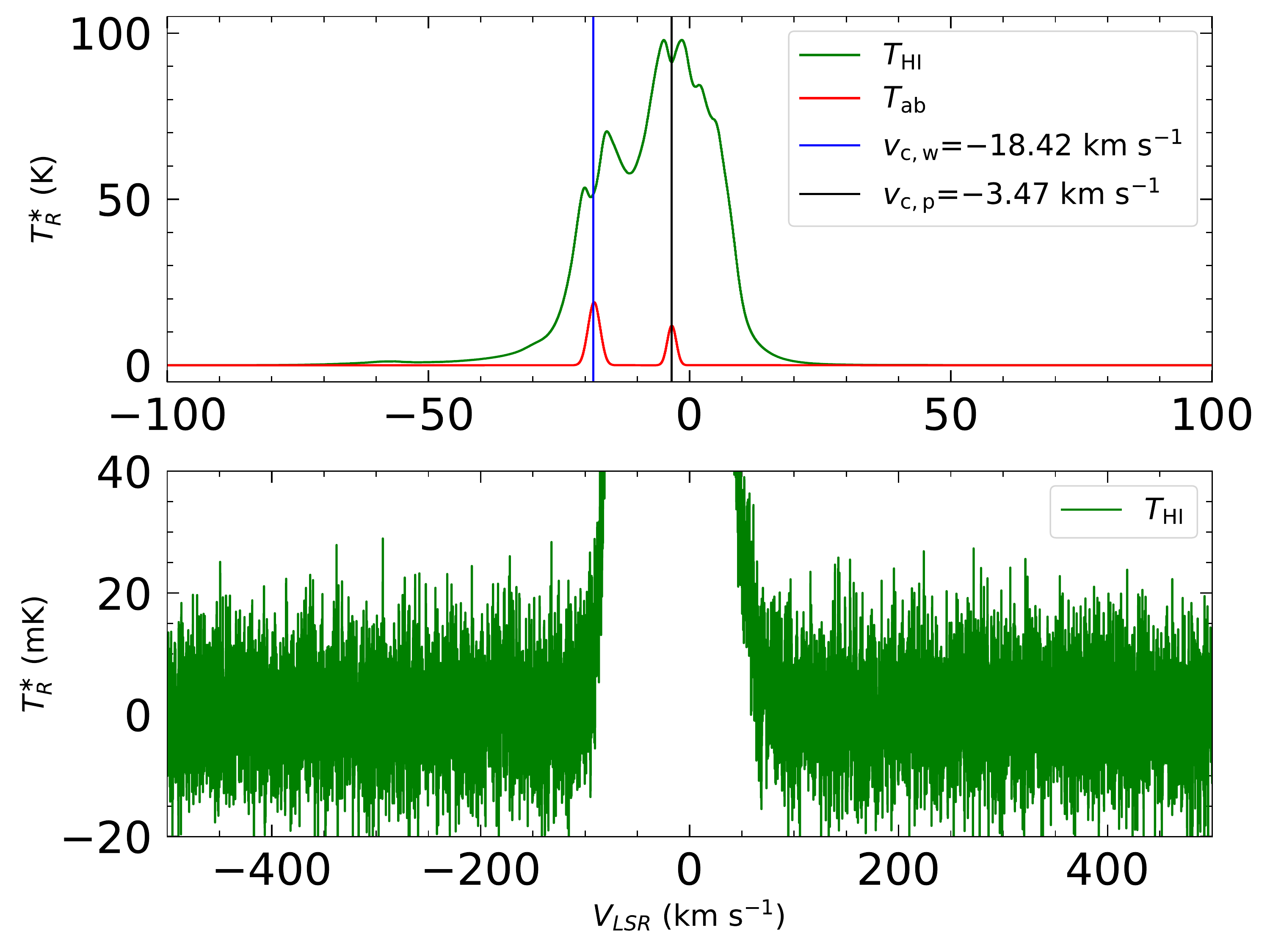}}
	\subfigure[Beam 2]{\includegraphics[height=0.245\textwidth]{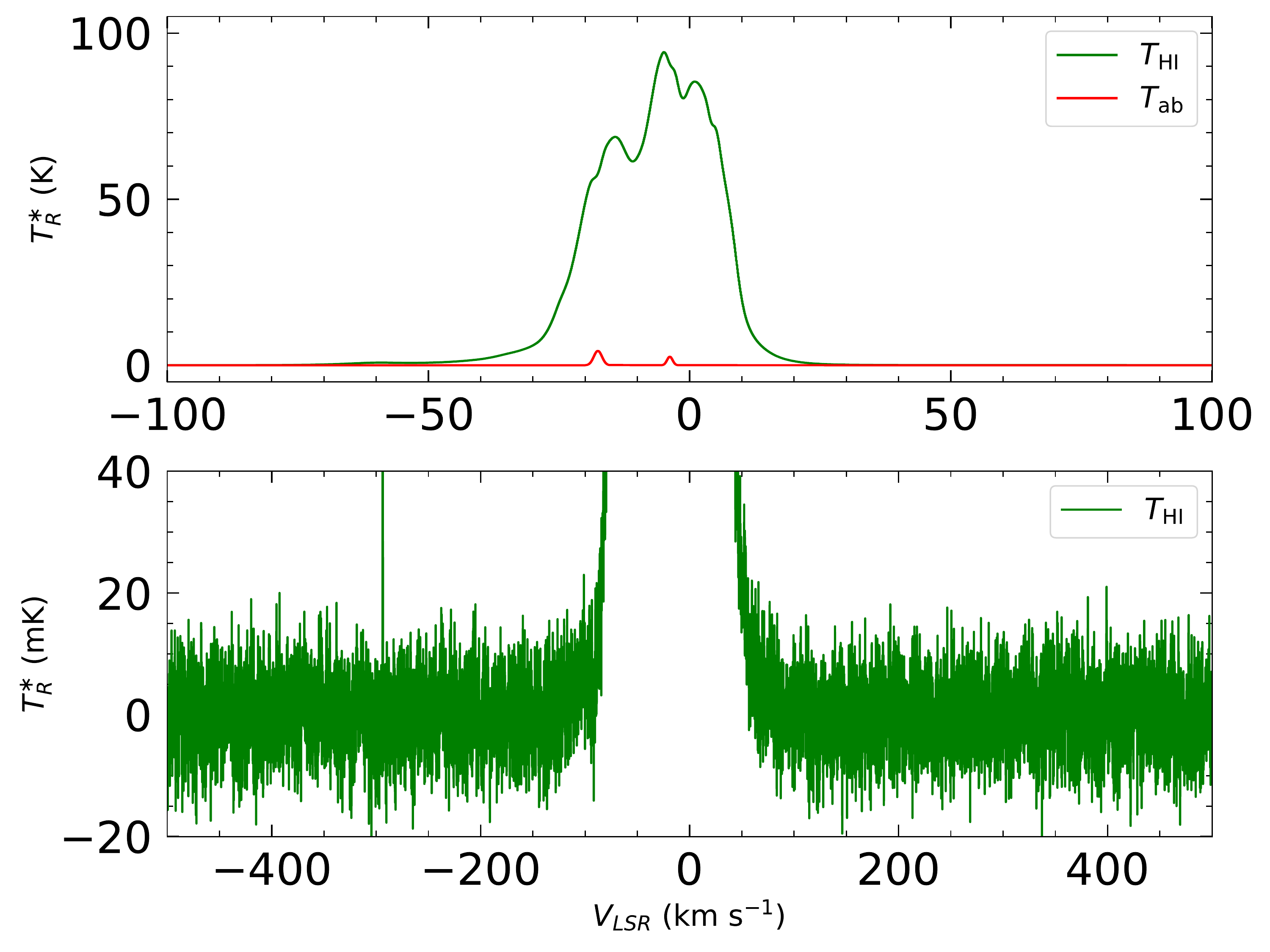}}
	\subfigure[Beam 3]{\includegraphics[height=0.245\textwidth]{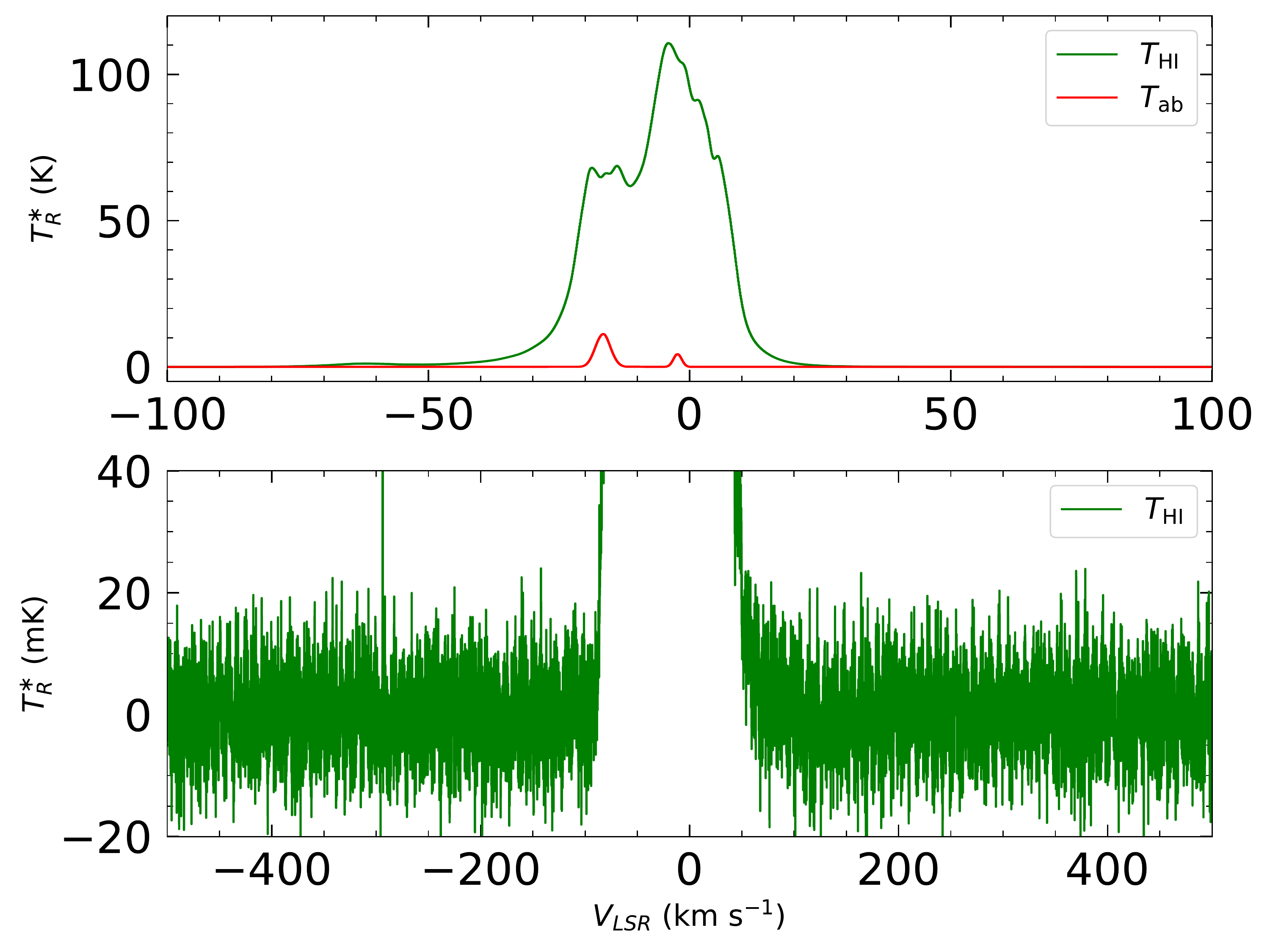}}
	\subfigure[Beam 4]{\includegraphics[height=0.245\textwidth]{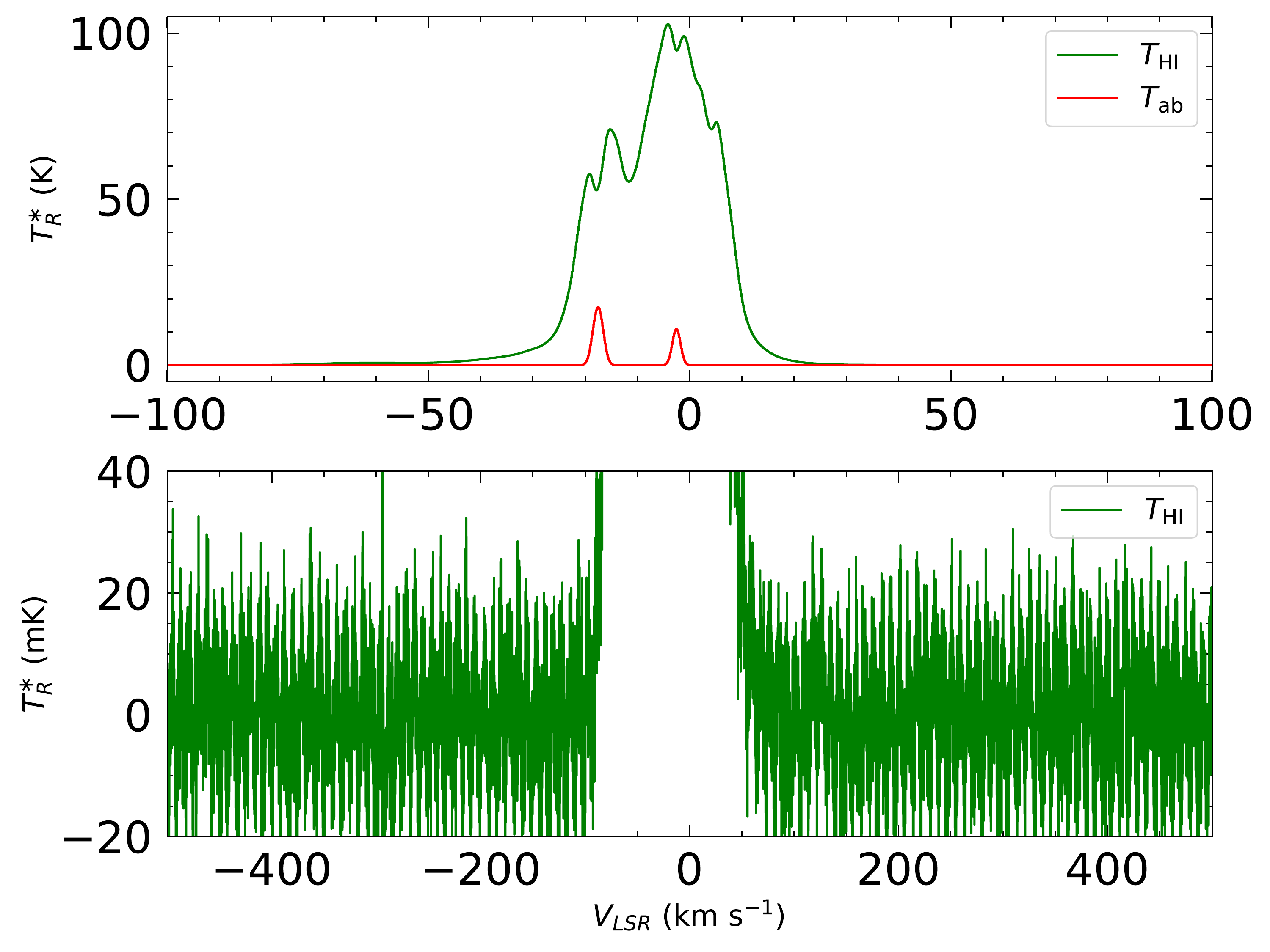}}
	\subfigure[Beam 5]{\includegraphics[height=0.245\textwidth]{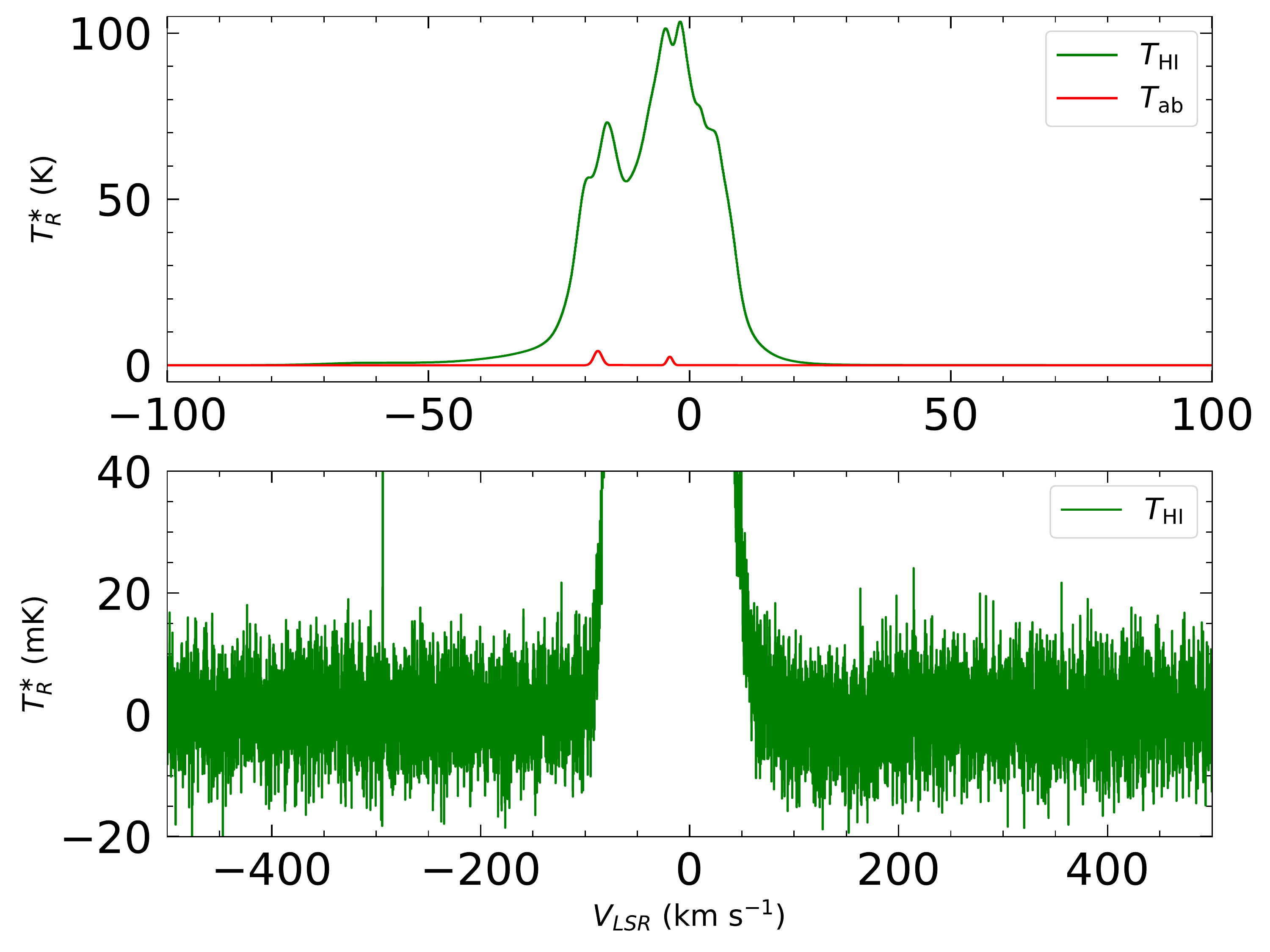}}
	\subfigure[Beam 6]{\includegraphics[height=0.245\textwidth]{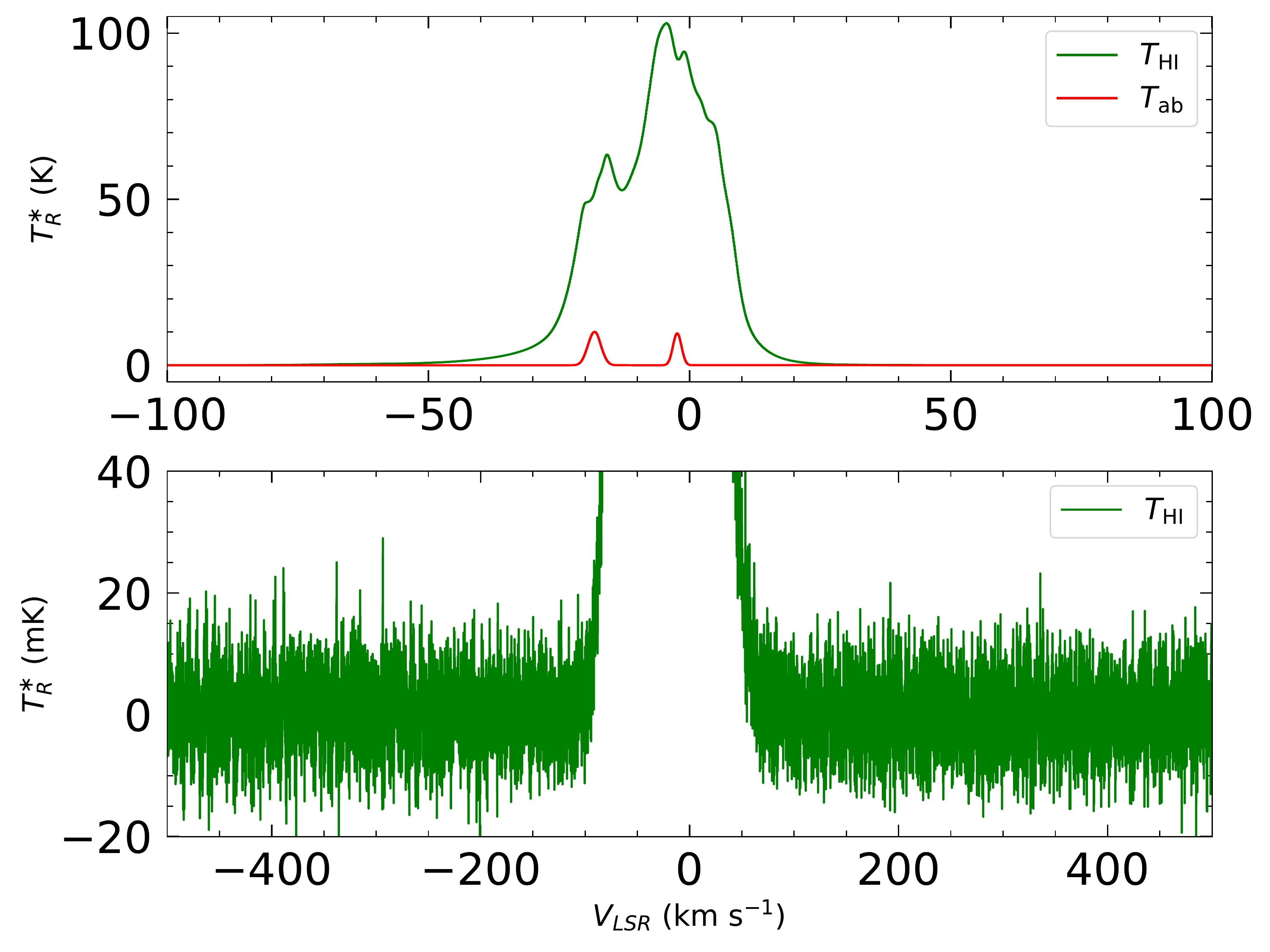}}
	\subfigure[Beam 7]{\includegraphics[height=0.245\textwidth]{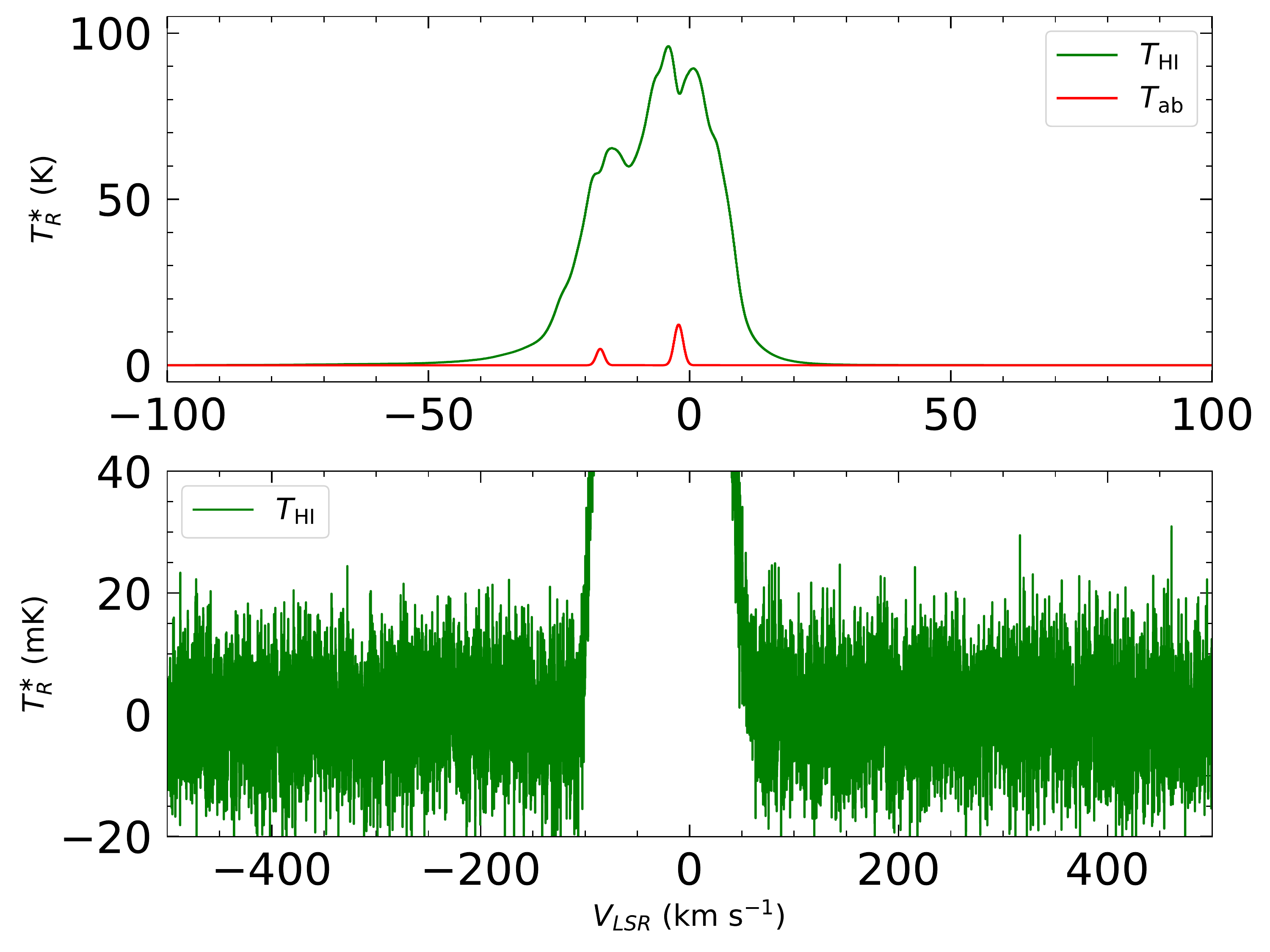}}	
	\caption{Top: spectra of the central seven beams (green lines) and the HINSA ($T_{\mathrm{ab}}$, red lines) obtained in \cite{Li+2022b}. Bottom: the flat baseline. $v_\mathrm{{c,w}}$ represents the central velocity of the HINSA feature (i.e., at $-$18.42 km s$^{-1}$) associated with the H~\textsc{i} wind (see below), and $v_\mathrm{{c,p}}$ represents the central velocity of another HINSA feature (i.e., at $-$3.47 km s$^{-1}$) in Beam 1.}
	\label{figure-noise}
\end{figure*}

Observations toward G176.51+00.20 were conducted on August 19th and 20th, 2021. The observational mode was 19-beam tracking, and the total integration time was 335 minutes with a sampling rate of one second. A new algorithm is used to obtain a flat baseline. In this algorithm, there are three main steps: (1) a polynomial fitting over the velocity range of [$-7000$, 7000] km s$^{-1}$; (2) removing the standing waves by using fast Fourier transforms; (3) calibrating the unsmooth parts in the baseline by extreme envelope curves \citep[for more details of the pipeline used to obtain the flat baseline, see][]{Liudj+2022}. For Beams 2, 3, 5, 6, and 8--19, the spectra of two linear polarizations (i.e., XX and YY) were dealt with separately and then averaged. However, because of the bad baseline, only the YY polarization spectrum from Beam 1 and the XX spectra from Beams 4 and 7 were retained. 
The mean rms noise of the nineteen spectra is $\sim$ 7 mK @ 0.1 km s$^{-1}$ (see the rms noise of each spectrum in Table \ref{tab:sources}). Figure \ref{figure-positions} presents the positions of the central seven beams superposed on CO molecular maps, and Figure \ref{figure-noise} shows the spectrum of the central seven beams and the H~\textsc{i} narrow-line self-absorption lines (HINSA, $T_{\mathrm{ab}}$) in each beam \citep[see][]{Li+2022b}.

\begin{deluxetable*}{cccc|cccc}
	\tabletypesize{\footnotesize}
	\tablecaption{List of the Positions and Rms Noise of the Nineteen Beams\label{tab:sources}}	
	\tablehead{
		\specialrule{0em}{1pt}{5pt}
		\colhead{Index}	&	\colhead{R.A.}	&	\colhead{Decl.} & \colhead{Rms} \vline & \colhead{Index}	&	\colhead{R.A.}	&	\colhead{Decl.} & \colhead{Rms} \\
		\colhead{} & \colhead{(J2000)} & \colhead{(J2000)} & \colhead{(mK)} \vline & \colhead{} & \colhead{(J2000)} & \colhead{(J2000)} & \colhead{(mK)}
	}
	\startdata
	1 & 05$^{\mathrm{h}}$37$^{\mathrm{m}}$53$^{\mathrm{s}}$ &  31$^{\circ}$59$'$58$''$ & 8.0 & 2 & 05$^{\mathrm{h}}$38$^{\mathrm{m}}$20$^{\mathrm{s}}$ &  32$^{\circ}$00$'$00$''$ & 6.1 \\		
	3 & 05$^{\mathrm{h}}$38$^{\mathrm{m}}$07$^{\mathrm{s}}$ &  31$^{\circ}$55$'$01$''$ & 6.8 & 4 & 05$^{\mathrm{h}}$37$^{\mathrm{m}}$39$^{\mathrm{s}}$ &  31$^{\circ}$54$'$59$''$ & 11.2 \\	
	5 & 05$^{\mathrm{h}}$37$^{\mathrm{m}}$26$^{\mathrm{s}}$ &  31$^{\circ}$59$'$57$''$ & 6.1 & 6 & 05$^{\mathrm{h}}$37$^{\mathrm{m}}$39$^{\mathrm{s}}$ &  32$^{\circ}$04$'$56$''$ & 6.2 \\ 
	7 & 05$^{\mathrm{h}}$38$^{\mathrm{m}}$06$^{\mathrm{s}}$ &  32$^{\circ}$04$'$57$''$ & 7.7 & 8 & 05$^{\mathrm{h}}$38$^{\mathrm{m}}$47$^{\mathrm{s}}$ &  32$^{\circ}$00$'$01$''$ & 8.8 \\ 
	9 & 05$^{\mathrm{h}}$38$^{\mathrm{m}}$34$^{\mathrm{s}}$ &  31$^{\circ}$55$'$02$''$ & 6.5 & 10 & 05$^{\mathrm{h}}$38$^{\mathrm{m}}$20$^{\mathrm{s}}$ &  31$^{\circ}$50$'$03$''$ & 7.7 \\ 
	11 & 05$^{\mathrm{h}}$37$^{\mathrm{m}}$53$^{\mathrm{s}}$ &  31$^{\circ}$50$'$02$''$ & 6.3 & 12 & 05$^{\mathrm{h}}$37$^{\mathrm{m}}$26$^{\mathrm{s}}$ &  31$^{\circ}$50$'$00$''$ & 5.9 \\ 
	13 & 05$^{\mathrm{h}}$37$^{\mathrm{m}}$12$^{\mathrm{s}}$ &  31$^{\circ}$54$'$58$''$ & 6.6 & 14 & 05$^{\mathrm{h}}$36$^{\mathrm{m}}$59$^{\mathrm{s}}$ &  31$^{\circ}$59$'$55$''$ & 6.9 \\ 
	15 & 05$^{\mathrm{h}}$37$^{\mathrm{m}}$12$^{\mathrm{s}}$ &  32$^{\circ}$04$'$54$''$ & 6.3 & 16 & 05$^{\mathrm{h}}$37$^{\mathrm{m}}$26$^{\mathrm{s}}$ &  32$^{\circ}$09$'$53$''$ & 6.9 \\  
	17 & 05$^{\mathrm{h}}$37$^{\mathrm{m}}$53$^{\mathrm{s}}$ &  32$^{\circ}$09$'$55$''$ & 6.1 & 18 & 05$^{\mathrm{h}}$38$^{\mathrm{m}}$20$^{\mathrm{s}}$ &  32$^{\circ}$09$'$56$''$ & 6.9 \\   
	19 & 05$^{\mathrm{h}}$38$^{\mathrm{m}}$33$^{\mathrm{s}}$ &  32$^{\circ}$04$'$58$''$ & 6.7          
	\enddata
\end{deluxetable*}

\section{Data Analysis and Results}\label{sec-HI-flow}

\subsection{Identification of the H~\textsc{i} Wind}\label{sec-HI-flow-identification}

We defined the excess brightness temperature (EBT, hereafter) as the spectrum resulting from the subtraction of the average spectrum of the off-positions from the spectrum of the on-position (i.e., the target beam). Similar to the work of \citet{Lizano+1988}, the six beams located immediately around the target beam (i.e., which acted as the on-position) were used as the off-positions. The baseline of the EBT was flat, i.e., the rms noise was $\sim$ 1.8--2.3 mK @ 2.0 km s$^{-1}$ (i.e., smooth over twenty channels), except for Beams 3 and 4, whose rms noises were, respectively, 3.3 and 6.1 mK @ 2.0 km s$^{-1}$. We also defined the standard deviation (STD) of EBT (SEBT) to evaluate the variation among the spectrum from different off-positions. For instance, for Beam 1, we first separately calculated the EBT of Beam 1 relative to its six surrounding beams, and then computed the STD of these six EBTs, which is the SEBT of Beam 1.

\begin{figure*}[!htb]
	\centering
	\subfigure[Beam 1]{\includegraphics[height=0.21\textwidth]{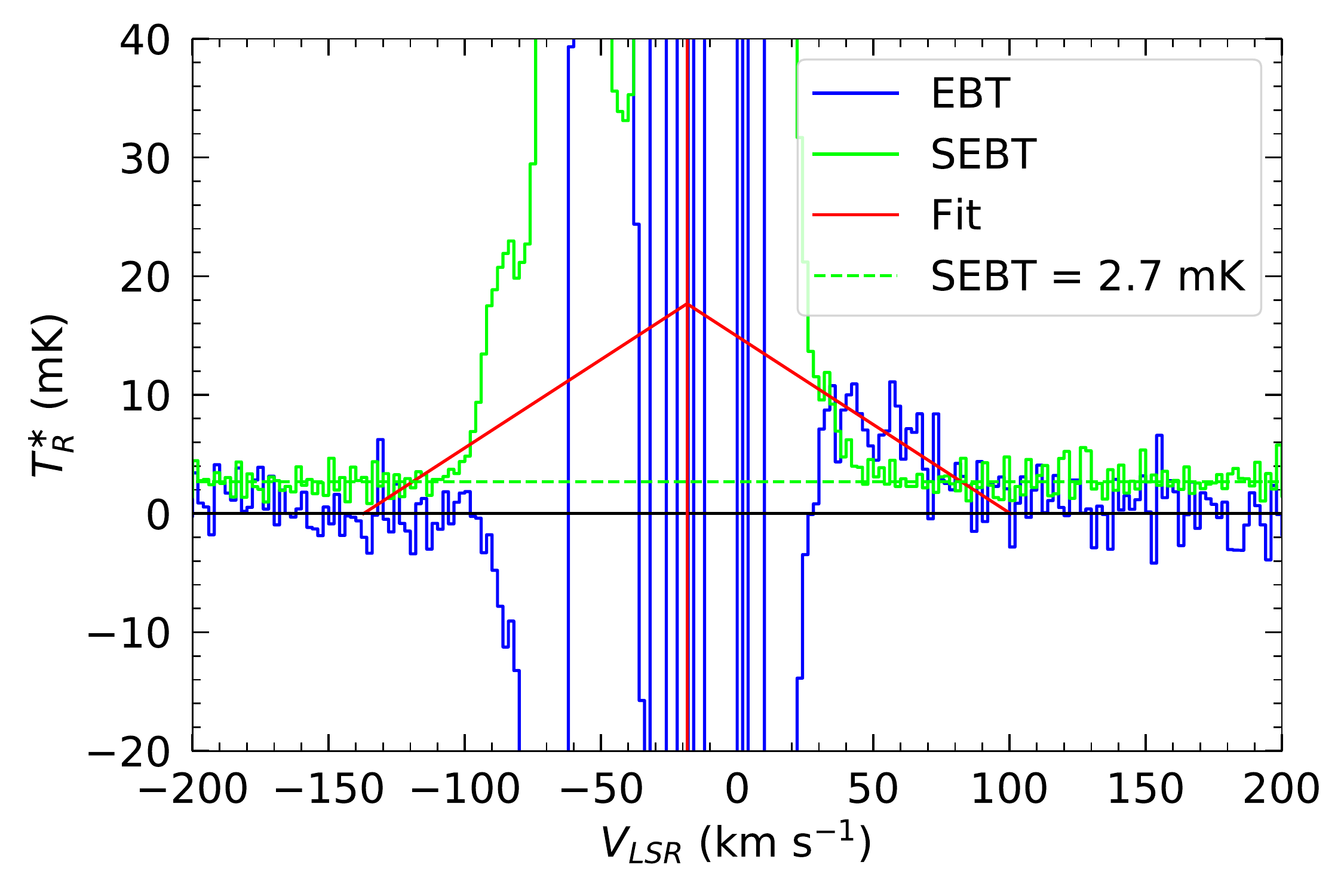}}
	\subfigure[Beam 2]{\includegraphics[height=0.21\textwidth]{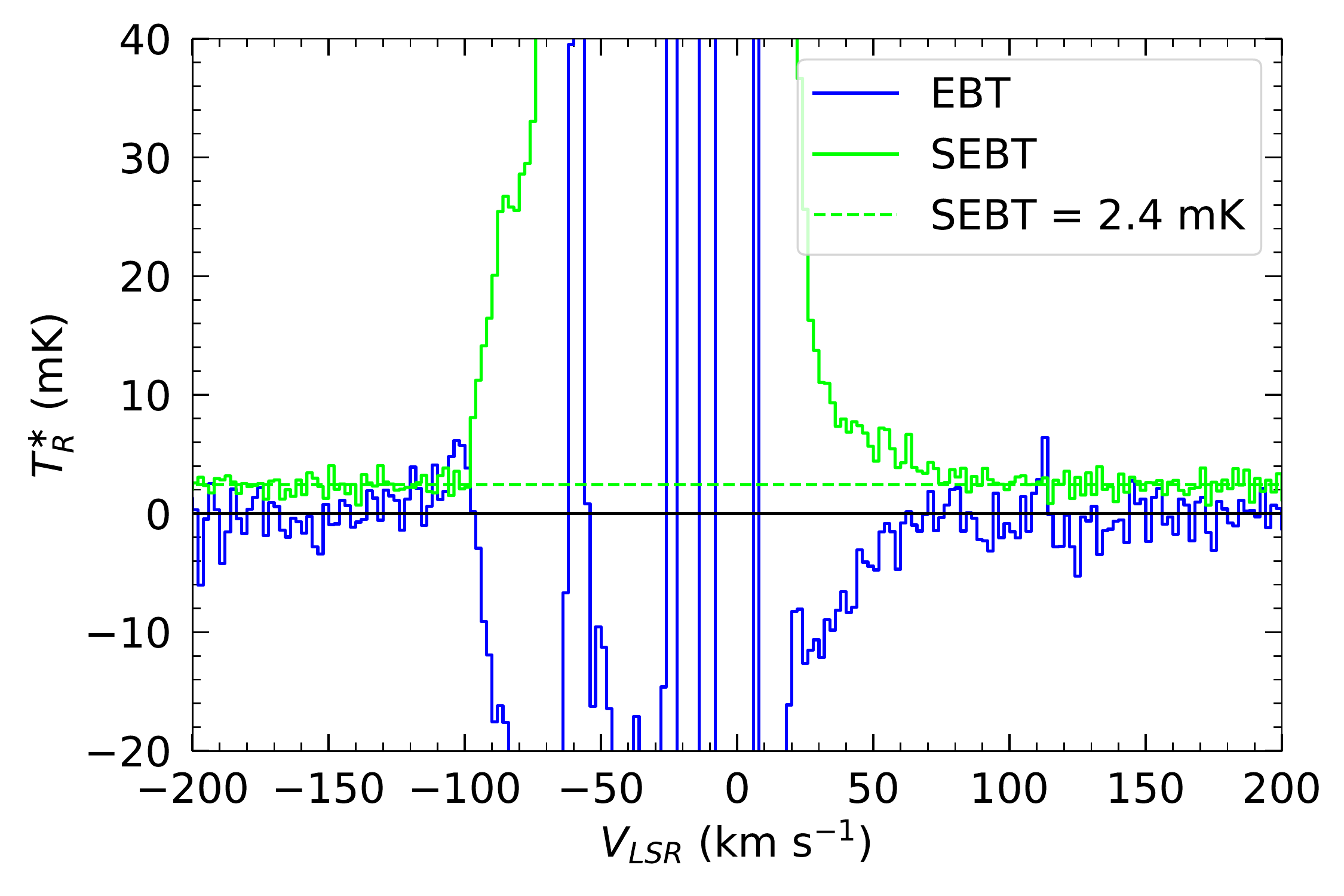}}
	\subfigure[Beam 3]{\includegraphics[height=0.21\textwidth]{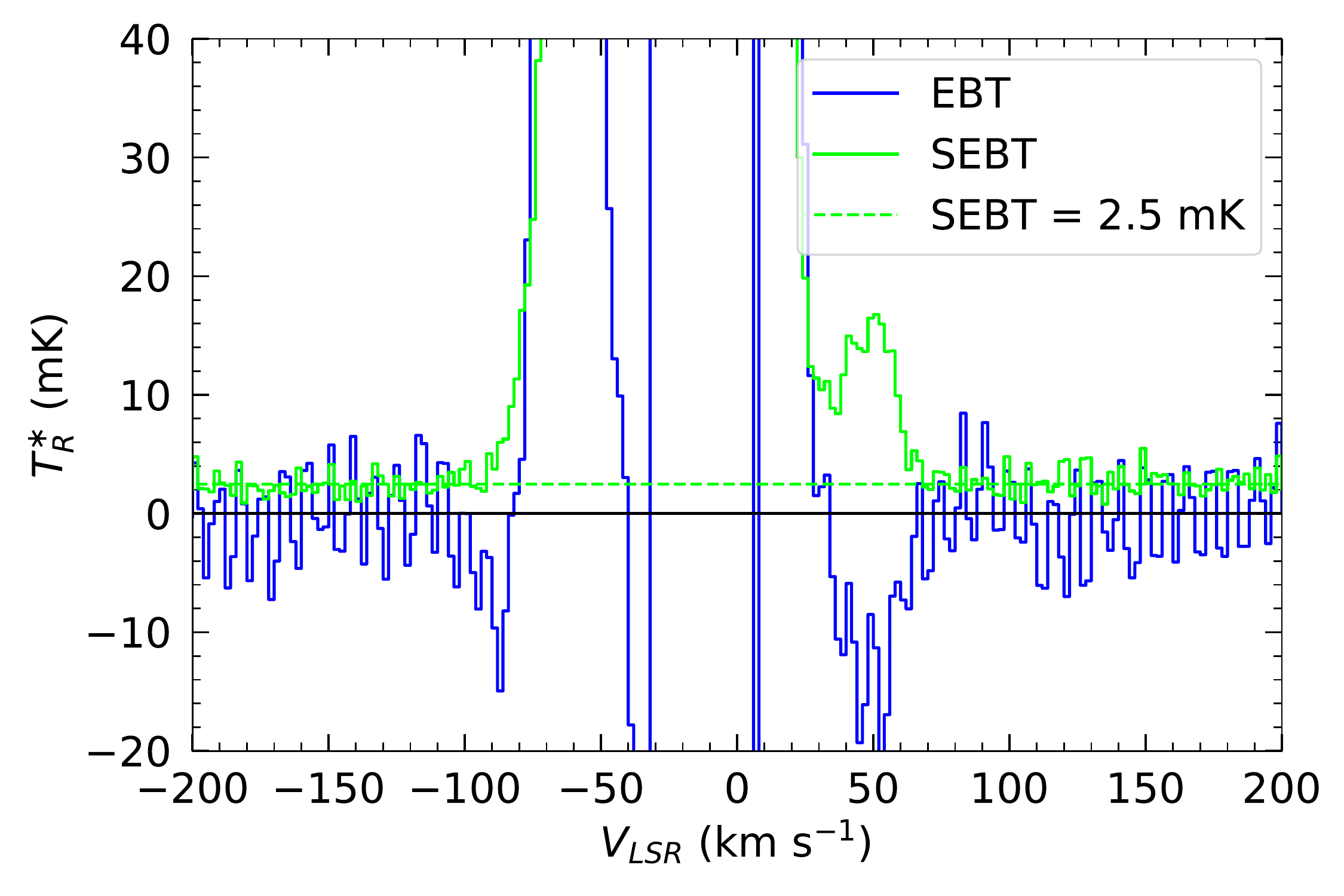}}
	\subfigure[Beam 4]{\includegraphics[height=0.21\textwidth]{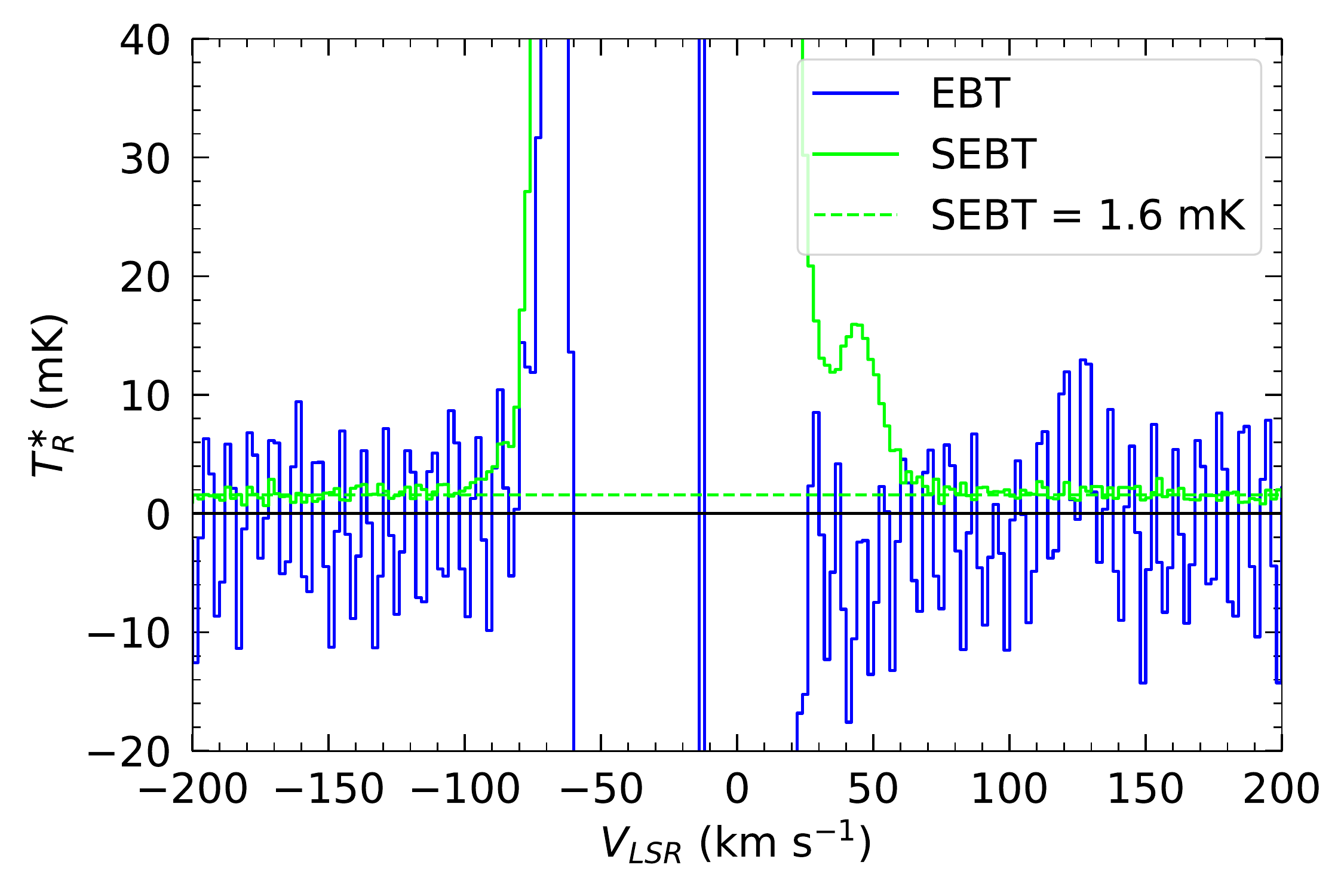}}
	\subfigure[Beam 5]{\includegraphics[height=0.21\textwidth]{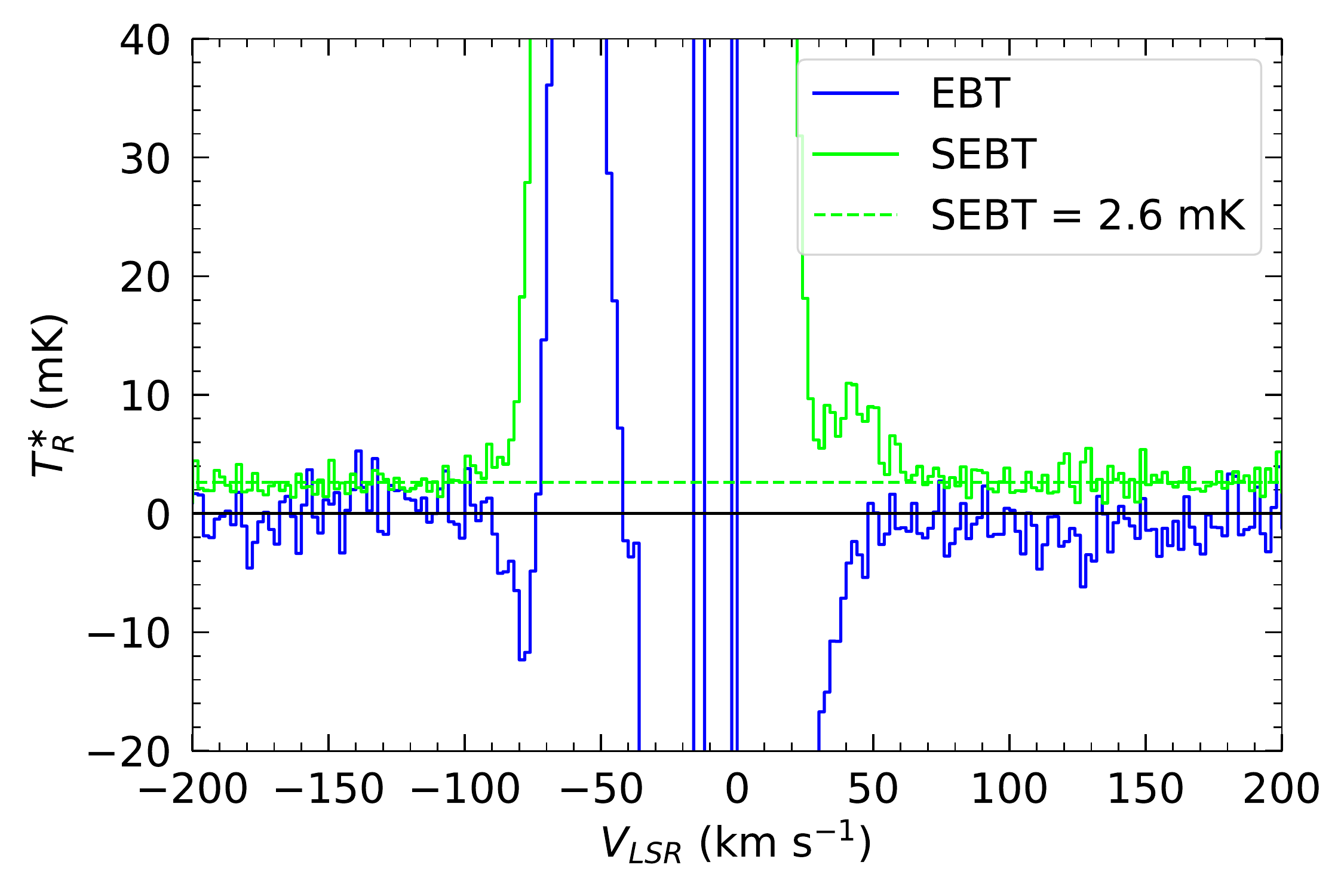}}
	\subfigure[Beam 6]{\includegraphics[height=0.21\textwidth]{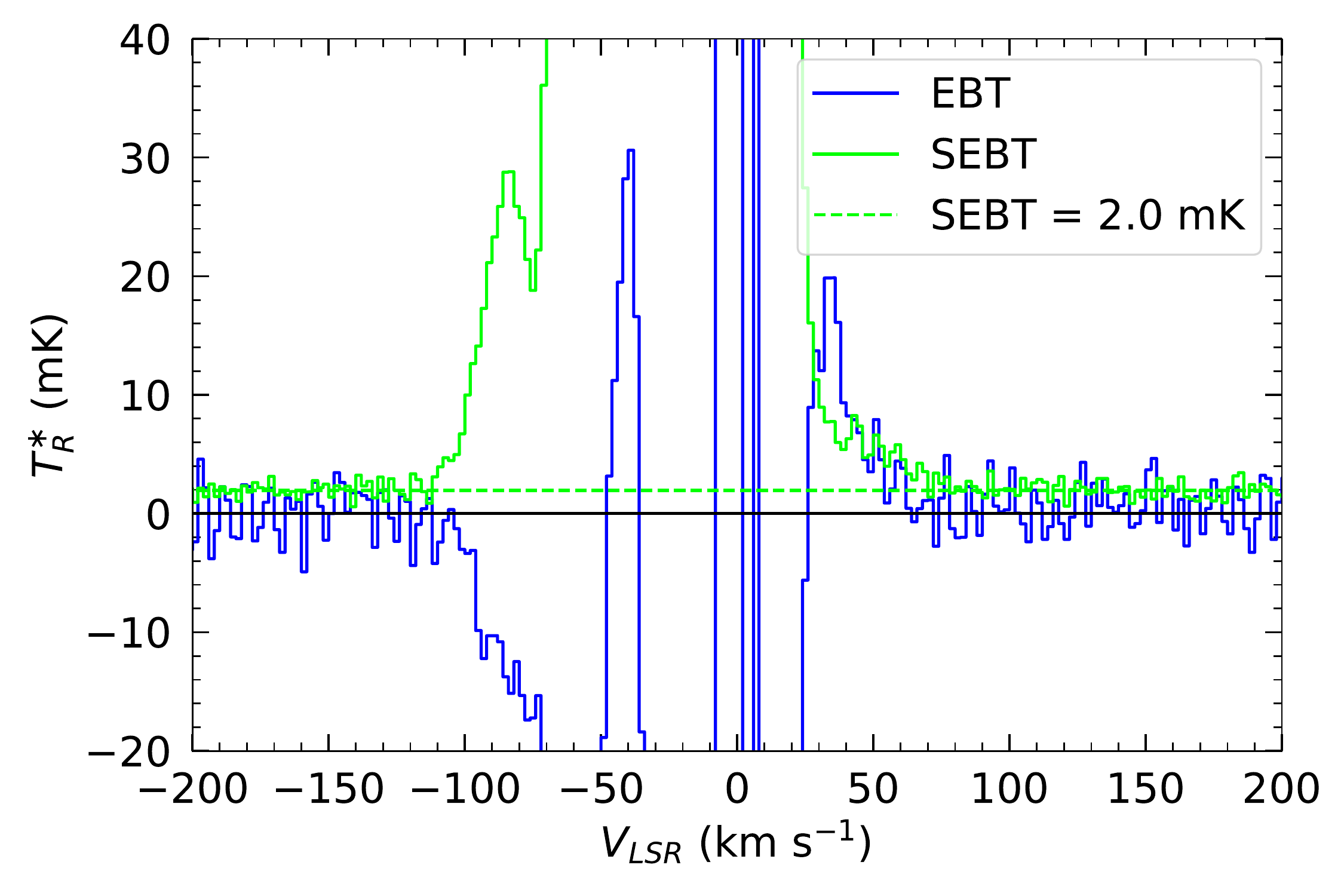}}
	\subfigure[Beam 7]{\includegraphics[height=0.21\textwidth]{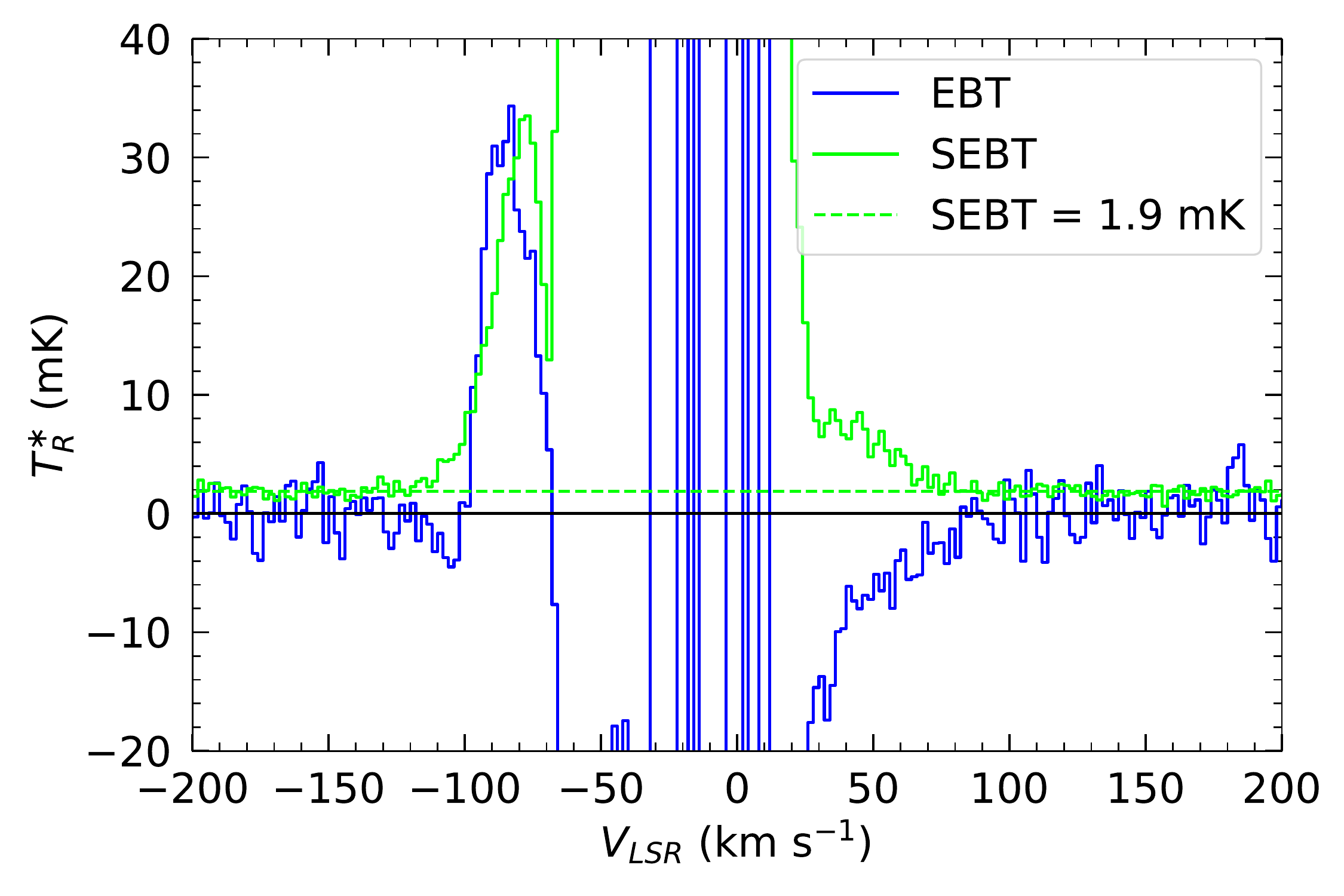}}	
	\caption{EBT (blue lines) and SEBT (green lines) of Beams 1--7 (smoothed over twenty channels, blue lines). For the spectrum in which the high-velocity line wing was present, the ``triangular fit'' is plotted as a red oblique line, where the line center (red vertical line) is taken as the velocity of the HINSA \citep[see Figure \ref{figure-noise} and ][]{Li+2022b}.}
	\label{Figure-HIflow-1}
\end{figure*}

\subsubsection{H~\textsc{i} Wind in Beam 1}

\begin{figure*}[!htb]
	\centering
	\subfigure[Lack Beam 7]{\includegraphics[height=0.3\textwidth]{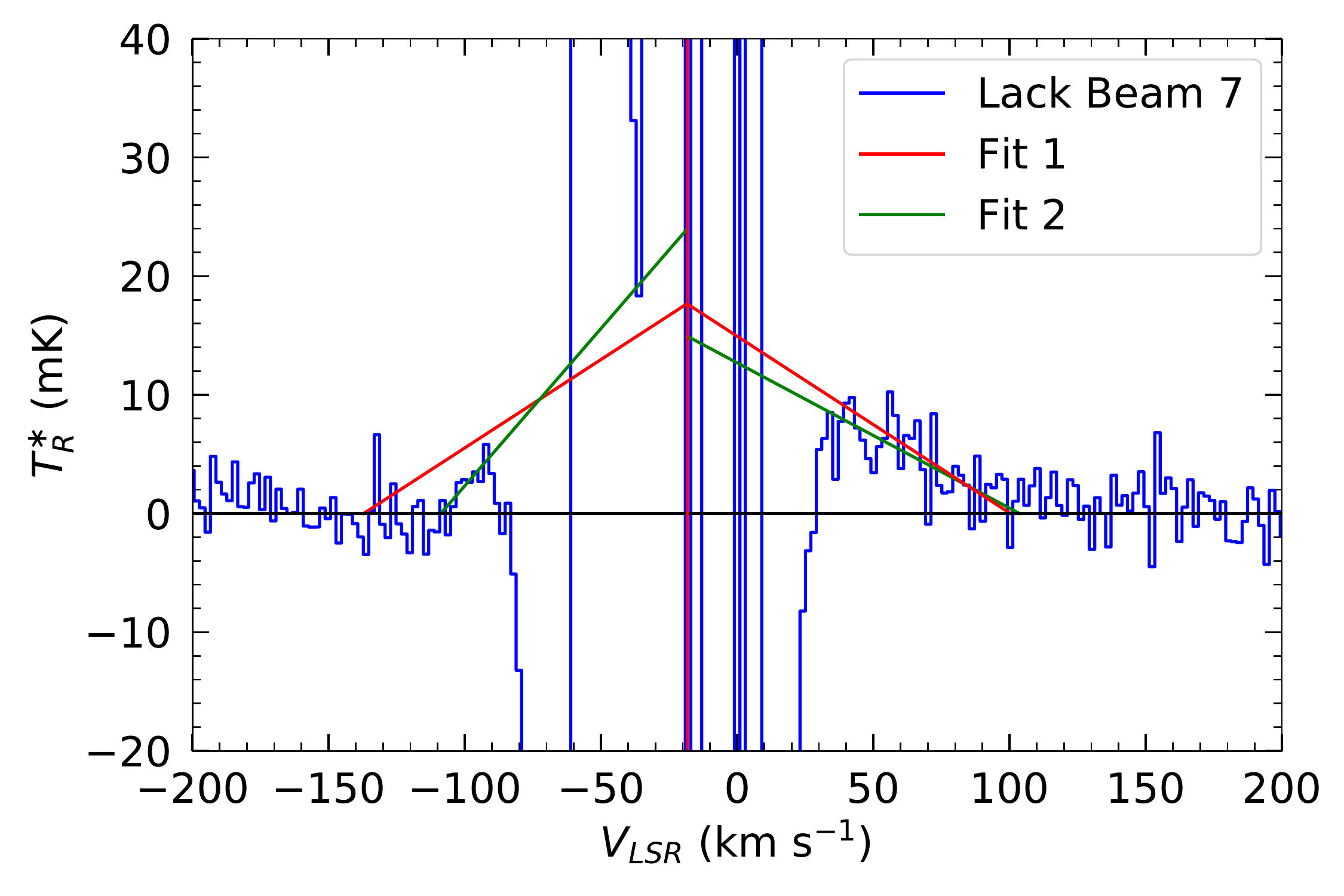}}
	\subfigure[W-E]{\includegraphics[height=0.3\textwidth]{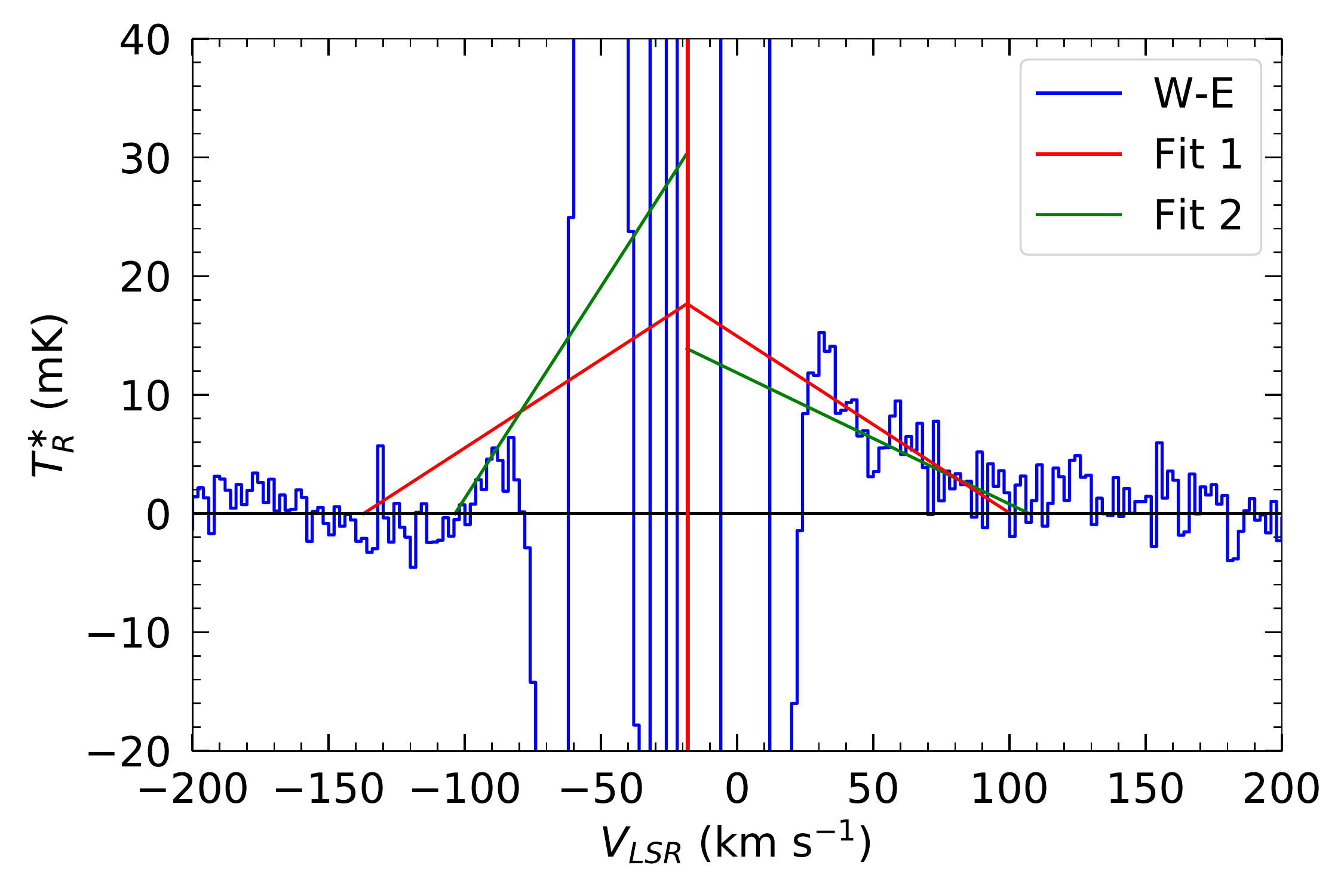}}
	\subfigure[NE-SW]{\includegraphics[height=0.3\textwidth]{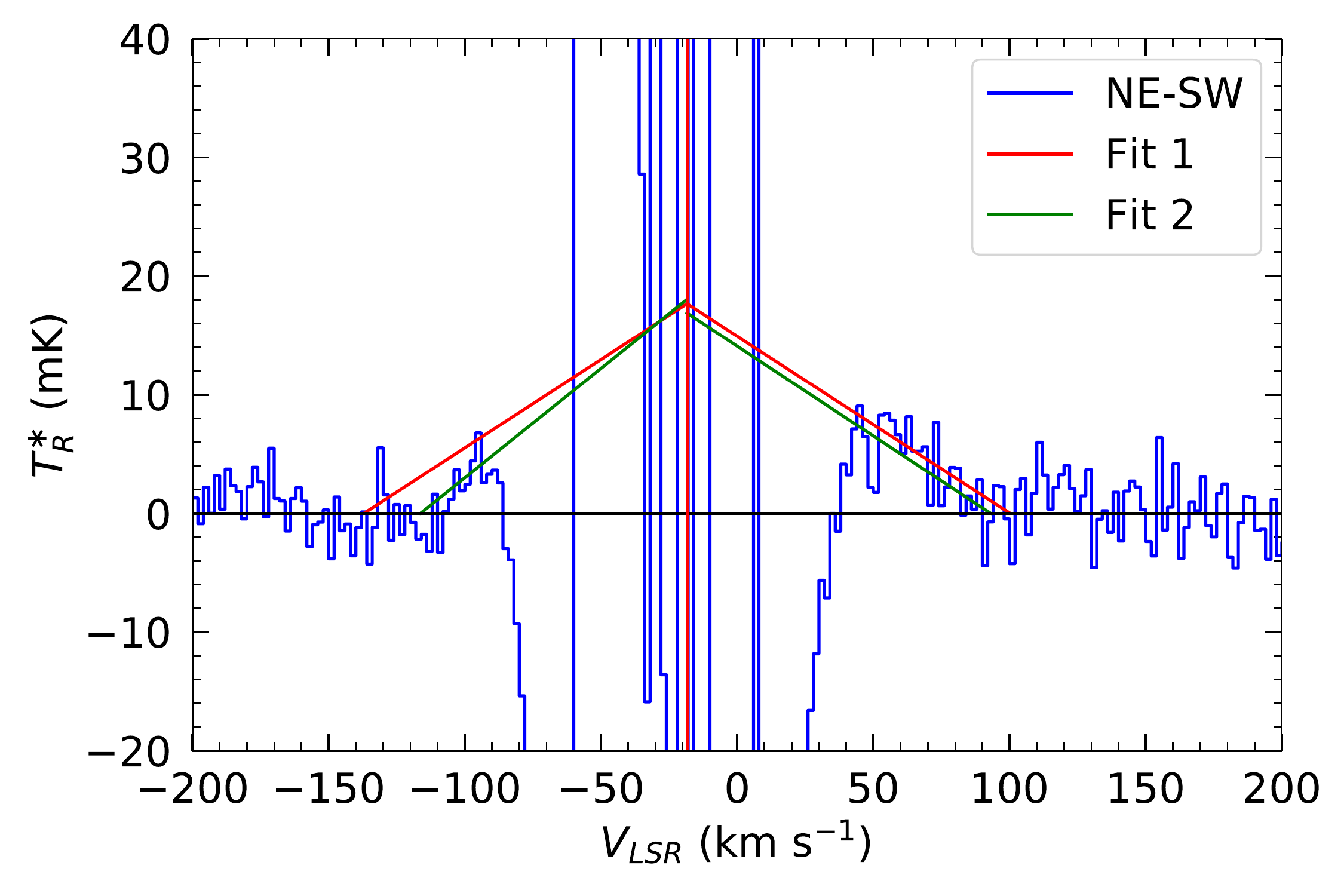}}
	\subfigure[NW-SE]{\includegraphics[height=0.3\textwidth]{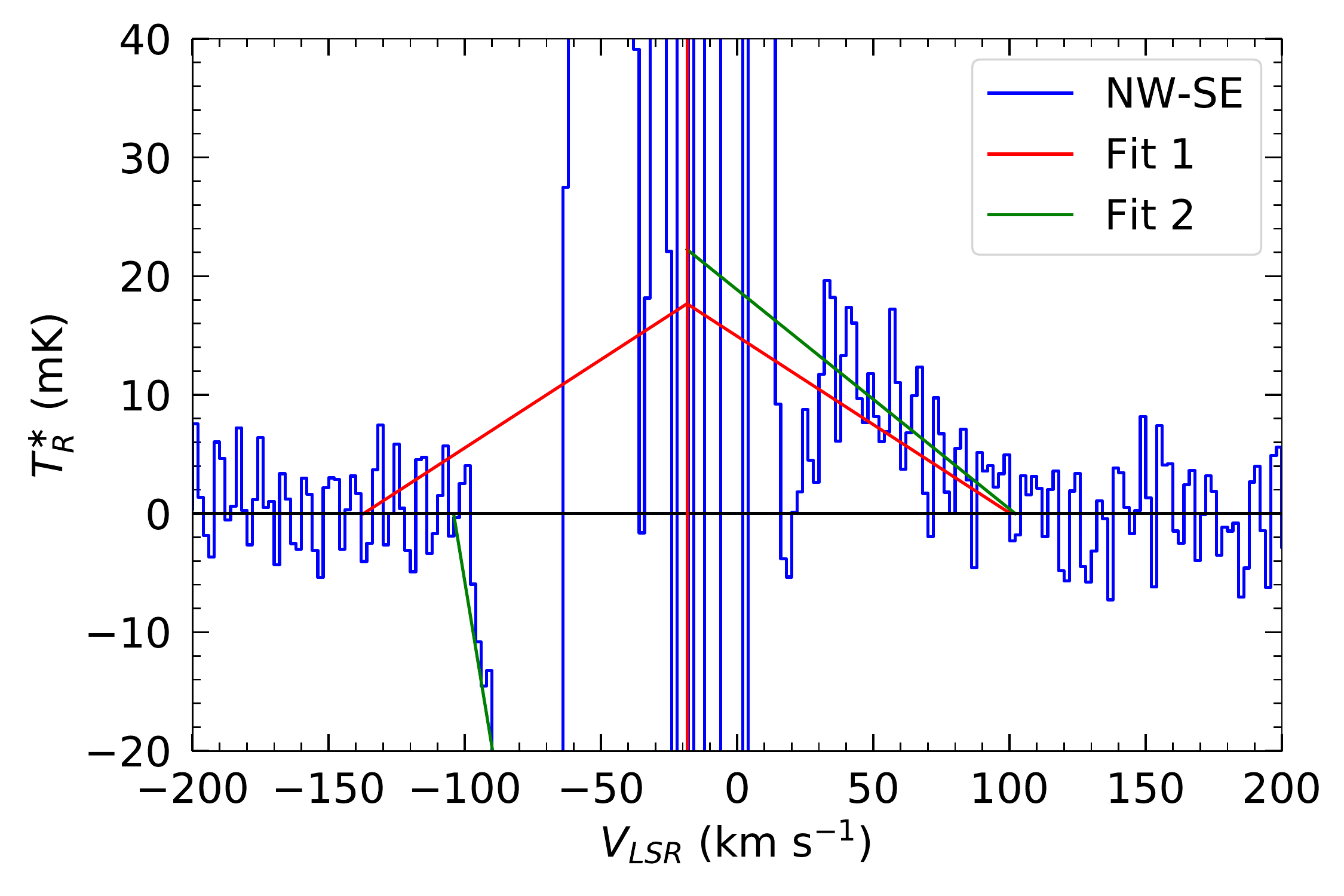}}	
	\caption{EBT obtained from using different beams as off-positions. The red lines (``Fit 1'') are the same as that in Figure \ref{Figure-HIflow-1}(a). The green lines (``Fit 2'') are the results of the ``triangular fit'' of the EBTs (fit separately for the blue and red lobes). Lack Beam 7: using Beams 2--6 as off-positions; W--E: taking Beams 2 and 5 as off-positions; NE--SW: using Beams 3 and 6 as off-positions; NW--SE: using Beams 4 and 7 as off-positions.}
	\label{Figure-HIflow-2}
\end{figure*}

From Figure \ref{Figure-HIflow-1}, it can be seen that Beam 1 presents an evident high-velocity red wing whose maximum velocity exceeds $\sim$ 120 km s$^{-1}$. The SEBT is also small in the high-velocity red wing, indicating this wing may well be true, rather than the changes or fluctuations of the emission from one beam area to another. This high-velocity red wing was crudely fitted by a ``triangular fit'' \citep[see e.g.,][]{Lizano+1988}. Assuming this high-velocity line wing is bipolar, the symmetric line of the high-velocity red wing with respect to the line center, $v_\mathrm{{c,w}}$, is also presented as a red oblique line (see the red isosceles triangle of the high-velocity bipolar wing in Figure \ref{Figure-HIflow-1}(a)). Therein, $v_\mathrm{{c,w}}$ is taken as the velocity of the corresponding HINSA \citep[see Figure \ref{figure-noise} and ][]{Li+2022b}, which is indicated as a red vertical line in Figure \ref{Figure-HIflow-1}(a). This symmetric line is referred to as the presumed high-velocity blue wing, although there is no blue wing present in Figure \ref{Figure-HIflow-1}(a). However, we find that the velocity of the bulge in Beam 7 (see Figure \ref{Figure-HIflow-1}(g)) overlaps with that of the presumed high-velocity blue wing in Beam 1. Therefore, a further analysis of the high-velocity line wing in Beam 1 was conducted where we took different beams as the off-positions (see Figure \ref{Figure-HIflow-2}).

When using Beams 2--7 as off-positions (see Figures \ref{Figure-HIflow-1}), we also took beams orientated along West--East (i.e., Beams 2 and 5), Northeast--Southwest (i.e., Beams 3 and 6), Northwest--Southeast (i.e., Beams 4 and 7), and also using Beams 2--6 as off-positions (see Figures \ref{Figure-HIflow-2}(a)--(d), respectively). These five cases are referred to as Normal, W--E, NE--SW, NW--SE, and Lack Beam 7. The high-velocity red wing remains basically unchanged no matter which beams are used as off-positions (see Figure \ref{Figure-HIflow-2} and quantitative comparison in Section \ref{sec-HI-flow-physical-properties}), and the high-velocity blue wing is also present as long as the off-positions do not contain Beam 7. Therefore, Beam 1 probably presents a high-velocity bipolar wing, but the high-velocity blue wing is contaminated by other components. In addition, there is a small pit between the ``triangular fit'' of the EBT for the cases of W--E, NE--SW, and NW--SE (see the left green oblique line in Figures \ref{Figure-HIflow-2}(a)--(c)) and the presumed high velocity blue wing (see red oblique line).  
This small pit likely causes the green oblique line to deviate from the red oblique line, which implies that this high-velocity bipolar wing is probably symmetrical; therefore, we identify it as a bipolar H~\textsc{i} wind. Because the blue lobe is contaminated by other components, and we presume that this H~\textsc{i} wind is a symmetrical bipolar H~\textsc{i} wind, we applied the fitting result of the red lobe to the blue lobe. 

\subsubsection{The EBT in Other Beams}

The bulge in the blue line wing of Beam 2 is, in fact, present in only one polarization spectrum (i.e., YY). Therefore, this blue line wing is highly likely a fake high-velocity line wing. Figure \ref{Figure-HIflow-1} also presents other high-velocity features, e.g., wings with negative intensities in right side of EBTs in Beams 2 and 7, and wings with positive intensities in right side of EBTs in Beam 6, etc. These high-velocity features are more likely to stem from the changes or fluctuations of the emission from one beam area to another, because these features are located where SEBTs are large.

\subsection{Physical Parameters of the H~\textit{\textsc{i}} Wind}\label{sec-HI-flow-physical-properties}

We calculated the mass, $M_{\mathrm{HI,wind}}$, momentum, $P_{\mathrm{HI,wind}}$, maximum velocity, $V_{\mathrm{max,wind}}$, and average velocity, $V_{\mathrm{mean,wind}}$, of the red lobe of the H~\textsc{i} wind (``triangular fit'') corresponding to different off-positions, i.e., the case of Normal (see red line in Figure \ref{Figure-HIflow-1}(a)), Lack Beam 7, W--E, NE--SW, and NW--SE (see green lines in Figures \ref{Figure-HIflow-2}(a)--(d)). The specific calculations without inclination correction are presented in Section \ref{sec-HIflow-mass-momentum}. The results show that all the values of $M_{\mathrm{HI,wind}}$, $P_{\mathrm{HI,wind}}$, $V_{\mathrm{max,wind}}$, and $V_{\mathrm{mean,wind}}$ are consistent with each other within the errors (all the errors presented in this work are 1$\sigma$ uncertainties) for all the five cases (see the physical values in Table \ref{tab:compare-off-spectrum}). This reinforces the conclusion that the red lobe of the H~\textsc{i} wind is real, and the case of Normal can reflect the physical properties of the red lobe of the H~\textsc{i} wind. Therefore, the calculations below are based on the case of Normal.

\begin{deluxetable*}{lcccc}
    \setlength\tabcolsep{18pt}
	\tablecaption{Comparison of the Physical Parameters Determined with Different Off-positions\label{tab:compare-off-spectrum}}	
	\tablehead{
		\colhead{Classification}	&	\colhead{$M_{\mathrm{HI,wind}}$}	&	\colhead{$P_{\mathrm{HI,wind}}$}	&	\colhead{$V_{\mathrm{max,wind}}$}	&	\colhead{$V_{\mathrm{mean,wind}}$}	\\
		\colhead{} &	\colhead{($M_{\odot}$)}	&	 \colhead{($M_{\odot}$ km s$^{-1}$)}	&	 \colhead{(km s$^{-1}$)}	&	\colhead{(km s$^{-1}$)}	}
	\startdata
		Normal	&	0.11 $\pm$ 0.02	&	4.03 $\pm$ 1.76 	&	120.6 $\pm$ 16.3	&	40.2 $\pm$ 5.4	\\
		Lack Beam 7 & 0.10 $\pm$ 0.02 & 3.48 $\pm$ 2.22 & 124.7 $\pm$ 20.7 & 41.6 $\pm$ 6.9 \\
		W-E & 0.09 $\pm$ 0.02 & 3.16 $\pm$ 3.56 & 129.7 $\pm$ 25.7 & 42.2 $\pm$ 8.6 \\  
		NE-SW & 0.10 $\pm$ 0.02 & 3.33 $\pm$ 1.65 & 112.6 $\pm$ 16.2 & 37.5 $\pm$ 5.4 \\
		NW-SE & 0.14 $\pm$ 0.02 & 5.20 $\pm$ 2.26 & 122.2 $\pm$ 16.0 & 40.7 $\pm$ 5.3
	\enddata
\end{deluxetable*}

The ``triangular fit'' of the red lobe of the H~\textsc{i} wind produced $T_{\mathrm{wind}}(v) = (14.94 \pm 1.32) + (-0.15 \pm 0.02) \cdot v$. The maximum velocity, $V_{\mathrm{max,wind}}$, and the average velocity, $V_{\mathrm{mean,wind}}$, without inclination correction are, respectively, $120.6 \pm 16.3$ and $40.2 \pm 5.4$ km s$^{-1}$, where $V_{\mathrm{mean,wind}} = V_{\mathrm{max,wind}}/3$ for the triangular profile. $V_{\mathrm{max,wind}}$ (and also $V_{\mathrm{mean,wind}}$) are 0.7 and 0.8 times that in HH 7--11 and in L1551 \citep[both are located in low-mass star-forming regions; see][]{Lizano+1988, Giovanardi+1992}, respectively. The column density of the red lobe of the H~\textsc{i} wind, $N_{\mathrm{HI,wind}}$, is (1.94 $\pm$ 0.30) $\times$ 10$^{18}$ cm$^{-2}$ (see Table \ref{tab:HIflow-properties}). Assuming that the H~\textsc{i} wind occupies the same spatial location as the molecular outflows, and both of them are angularly unresolved \citep[e.g.,][]{Lizano+1988}, $N_{\mathrm{HI,wind}}$ is two orders of magnitude smaller than the column density of the outflow traced by $^{12}$CO, and three orders of magnitude smaller than that traced by $^{13}$CO, HCO$^+$, and CS \citep[see the column density of the molecular outflows in][]{Liu+2021}.

$^{13}$CO is proved to be a great tracer of the molecular column density, and the fractional HINSA abundance (where the molecular column density is traced by $^{13}$CO) is $\sim$ 1.1 $\times$ 10$^{-3}$ \citep{Li+2022b}. We also calculated the abundance of H~\textsc{i} in the H~\textsc{i} wind, $X_{\mathrm{HI,wind}}$, as:
\begin{equation}\label{equ:X-HI-flow}
	X_{\mathrm{HI,wind}} = \frac{N_{\mathrm{HI,wind}}}{N_{\mathrm{HI,wind}} + 2N_{\mathrm{H_2,outflow}}},
\end{equation}
where $N_{\mathrm{H_2,outflow}}$ is the column density of the molecular outflow \citep[that is traced by $^{12}$CO, $^{13}$CO, HCO$^+$, or CS; see][]{Liu+2021}. The results of $X_{\mathrm{HI,wind}}$ are listed in Table \ref{tab:HIflow-properties}. $X_{\mathrm{HI,wind}}$, determined by the molecular outflow traced by $^{13}$CO, is $\sim$ 1.0 $\times$ 10$^{-3}$ (only for the red lobe), which is consistent with the fractional HINSA abundance.

This consistency reveals an internal correlation between the H~\textsc{i} wind and HINSA. It also supports the conclusion of the mixture of the H~\textsc{i} wind and the molecular outflow because HINSA is probably mixed with the gas in the cold, well-shielded regions of the molecular clouds \citep[see][]{Li-Goldsmith2003, Li+2022b}. However, observations with higher resolution are required to confirm the correlation between the H~\textsc{i} wind and HINSA, and a larger sample is essential to rule out that the observed consistency here is only a coincidence.

\begin{deluxetable*}{lll}
    \tabletypesize{\footnotesize}
	\tablecaption{Physical Properties of the H~\textsc{i} Wind without Inclination Correction\label{tab:HIflow-properties}}

	\tablehead{
		\colhead{Quantity}	&	\colhead{Value (1$\sigma$ uncertainty)}	&	\colhead{Noting}	}
	\startdata
		$N_{\mathrm{HI,wind}}$ (cm$^{-2}$) 	&	(1.94 $\pm$ 0.30) $\times$ 10$^{18}$ & See Equation (\ref{NHI-flow-red-beam1})	\\
		$X_{\mathrm{HI,wind}}$ &  $\sim$ (4.6, 1.0, 0.4, 0.8) $\times$ 10$^{-3}$ & For $^{12}$CO, $^{13}$CO, HCO$^+$ and CS, See Equation (\ref{equ:X-HI-flow})  \\
		$M_{\mathrm{HI,wind}}$ ($M_{\odot}$) & 0.22 $\pm$ 0.03  & Combined two lobes, see Equation (\ref{MHI-flow-red-beam1}) \\
		$P_{\mathrm{HI,wind}}$ ($M_{\odot}$ km s$^{-1}$) & 8.05 $\pm$ 3.52  & Combined two lobes, see Equation (\ref{PHI-flow-red-beam1}) \\	
		$V_{\mathrm{max,wind}}$ (km s$^{-1}$) & 120.6 $\pm$ 16.3 & See Equation (\ref{Vmax-red-beam1}) \\
		$V_{\mathrm{mean,wind}}$ (km s$^{-1}$) & 40.2 $\pm$ 5.4  & $V_{\mathrm{max,wind}}/3$ \\
		$E_{\mathrm{HI,wind}}$ (erg) & (3.4 $\pm$ 1.5) $\times$ 10$^{46}$  & See Equation (\ref{EHI-flow-red-beam1}) \\	
		$\Delta M_{\ast, \mathrm{m}}$ ($M_{\odot}$) & $\sim$ 0.43, $\sim$ 0.61, $\sim$ 3.68, $\sim$ 1.53  & For $^{12}$CO, $^{13}$CO, HCO$^+$ and CS \\
		$\Delta M_{\ast, \mathrm{ame}}$ ($M_{\odot}$) & 1.3, $-$5.8, 10.5, 6.5, 9.0, 2.7 & Respectively relative to Beams 2--7 \\
		$l_{\mathrm{HI,wind}}$ (pc) & 0.8 & See $l_{\mathrm{outflow}}$ in \citet{Liu+2021} \\
		$t_{\mathrm{a}}$ (yr) & (1.9 $\pm$ 0.2) $\times$ 10$^4$	& $ l_{\mathrm{HI,wind}}/V_{\mathrm{mean,wind}}$\\
		$\dot{M}_{\mathrm{a}}$ ($M_{\odot}$ yr$^{-1}$) & (12.2 $\pm$ 3.6) $\times$ 10$^{-6}$	& $ M_{\mathrm{HI,wind}}/t_{\mathrm{a}}$\\
		$t$ (yr) & (5.3 $\pm$ 3.5) $\times$ 10$^5$	& $ \Delta M_{\ast, \mathrm{a}}/\dot{M}_{\mathrm{a}}$\\
		$L_{\mathrm{HI,wind}}$ ($L_{\odot}$) & 16.0 $\pm$ 10.1 & $\displaystyle \frac{1}{2} \dot{M}_{\mathrm{a}} V^2_{\mathrm{max,wind}}$
	\enddata
\end{deluxetable*}

We also calculated the mass, $M_{\mathrm{HI,wind}}$, momentum, $P_{\mathrm{HI,wind}}$, and kinetic energy, $E_{\mathrm{HI,wind}}$, of the H~\textsc{i} wind (combine the values of the two lobes) without inclination correction (see the calculations in Section \ref{sec-HIflow-mass-momentum} and the values in Table \ref{tab:HIflow-properties}). 
The mass and momentum over the overall flow history are larger than the values estimated above \citep[see][]{Lizano+1988}. The values of $M_{\mathrm{HI,wind}}$, $P_{\mathrm{HI,wind}}$, and $E_{\mathrm{HI,wind}}$ are larger than that in HH 7--11 \citep[see][]{Lizano+1988, Giovanardi+1992} by a factor of $\sim$ 7, $\sim$ 5, and $\sim 4$, respectively, and larger than that in L1551 \citep[see][]{Giovanardi+1992} by a factor of $\sim$ 26, $\sim$ 20, and $\sim$ 6, respectively. 

\section{Discussion}\label{sec-discussion}

\subsection{Dynamical Comparison Between the H~\textsc{i} Wind and the Molecular Outflow}\label{sec-HI-flow-compare-with-molecular}

To determine whether the H~\textsc{i} wind is strong enough to drive the molecular outflow, analyses of two dimensions were conducted and are presented here; i.e., the mass loss of the excitation star and the lifetime of the H~\textsc{i} wind.

\subsubsection{Mass Loss of the Excitation Star}\label{sec-HI-flow-compare-with-molecular-mass-loss}

If we assume that the molecular outflow is entrained by the H~\textsc{i} wind originally ejected from the star, then the mass loss, $\Delta M_{\ast, \mathrm{m}}$, from the excitation star over the entire history of mass loss is $M_{\ast, \mathrm{m}} = P_{ \mathrm{m, outflow}}/V_{\mathrm{max,wind}}$ \citep[see][]{Lizano+1988}, where $P_{ \mathrm{m, outflow}}$ is the momentum of the molecular outflow \citep[see][]{Liu+2021}. Both the blue and red lobes of the molecular outflow are included in the calculation. The total $\Delta M_{\ast, \mathrm{m}}$, which combines contributions from the $^{12}$CO, $^{13}$CO, HCO$^+$, and CS outflows, is 6.26 $\pm$ 0.80 $M_{\odot}$ (see the contribution from each molecular outflow in Table \ref{tab:HIflow-properties}). Because \citet{Liu+2021} did not give the error of $P_{ \mathrm{m, outflow}}$, the error of $\Delta M_{\ast, \mathrm{m}}$ is only from $V_{\mathrm{max,wind}}$ and no error is presented in Table \ref{tab:HIflow-properties}. It is worth noting that the total $\Delta M_{\ast, \mathrm{m}}$ is probably overestimated because the values of $P_{\mathrm{m, outflow}}$ are corrected for the abundance of the corresponding tracer \citep[see][]{Liu+2021}.

\begin{figure*}[!htb]
	\centering
	\includegraphics[height=0.45\textwidth]{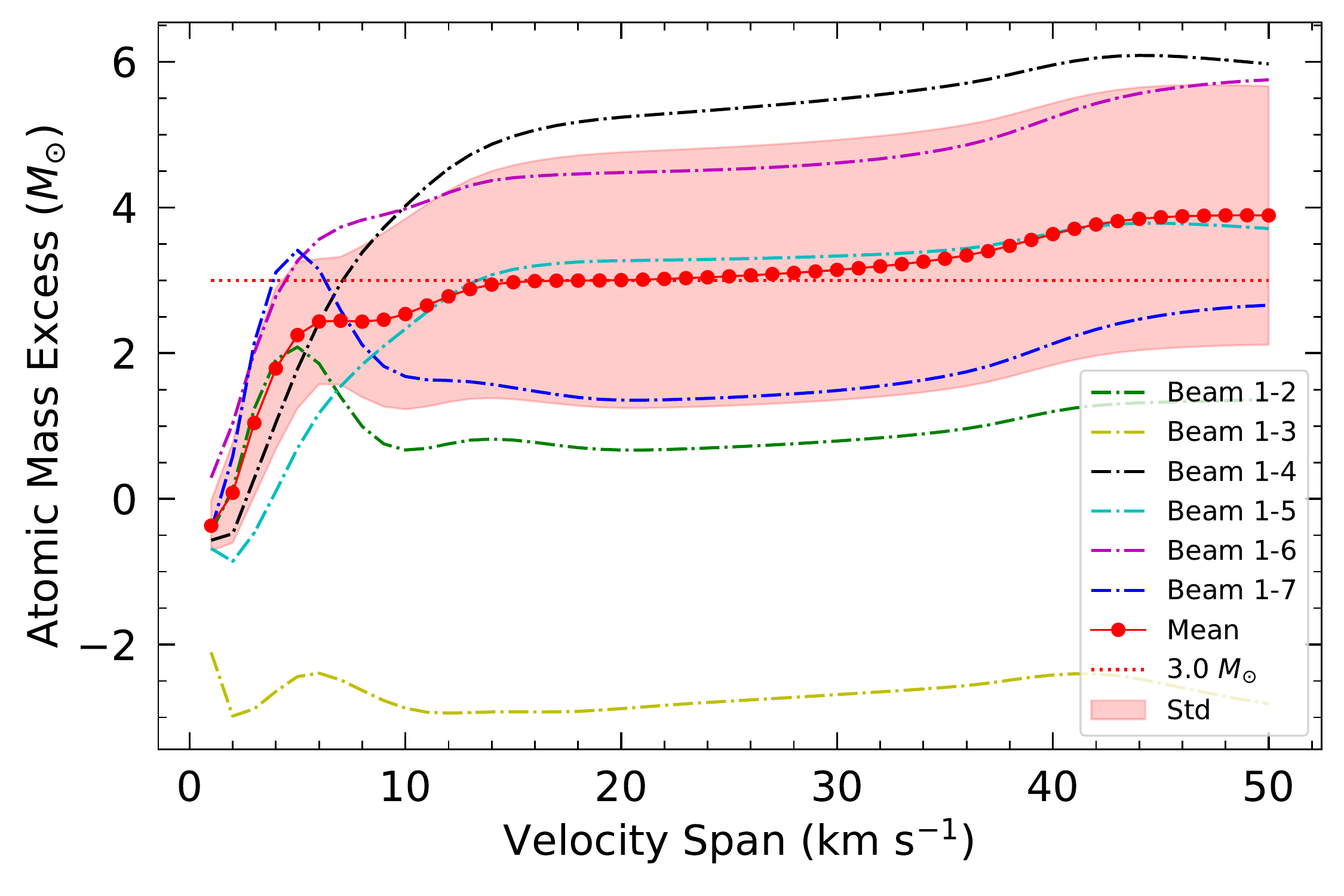}
	\caption{Atomic mass excesses, $\Delta M_{\ast, \mathrm{ame}}$, as a function of velocity spans, $v_{\mathrm{s}}$. ``Beam 1-$i$'' represent the atomic mass excess of Beam 1 relative to Beam $i$, ``Mean'' and ``Std'' present the mean and STD values of the five atomic mass excesses (i.e., except for that between Beams 1 and 3),  and the red dashed line that labeled as ``3.0 $M_{\odot}$'' indicates $\Delta M_{\ast, \mathrm{ame}} = 3.0$ $M_{\odot}$.}
	\label{fig-mass-excess}
\end{figure*}

Similar to the work in \citet{Lizano+1988}, we calculated the atomic mass excesses, $\Delta M_{\ast, \mathrm{ame}}$, of the line core of Beam 1 relative to the line core of the six beams tightly around Beam 1, assuming that the low-velocity material is associated with the source and fills the beams of FAST. We tested different velocity spans, $v_{\mathrm{s}}$, as line core (see Figure \ref{fig-mass-excess}), where $v_{\mathrm{s}} = v_{\mathrm{c,w}} - v$ and only $v < v_{\mathrm{c,w}}$ is considered because there is a strong component at $\sim$ $-$3.5 km s$^{-1}$. The increase of $\Delta M_{\ast, \mathrm{ame}}$ at $v_{\mathrm{s}} \sim$ 30--45 km s$^{-1}$ may be the results of the influence of the component at $\sim -57$ km s$^{-1}$ (see Figure \ref{figure-noise}). $\Delta M_{\ast, \mathrm{m}}$ is flat at $v_{\mathrm{s}} \sim$ 15--23 km s$^{-1}$, indicating that the influence of the component at $\sim$ $-$3.5 km s$^{-1}$ may be small. We set the velocity range of $v_\mathrm{{c,w}} - v \leq 20$ km s$^{-1}$ as the line core, corresponding to $v_{\mathrm{s}} \sim$ 20 km s$^{-1}$ (i.e., at the middle of the flat region). The atomic mass excesses of Beams 1 to its surrounding beams are systematically positive (except for those between Beams 1 and 3). In addition, such systematically positive of the atomic mass excesses is invalid for the case of Beams 2 and 4--7 relative to their surrounding beams. \footnote{For Beams 2 and 4--7, the mass excess relative to their six surrounding beams is not systematically positive. For instance, the atomic mass excesses of Beam 2 relative to its six surrounding beams are, respectively, $-0.4$, $-4.6$, $-3.2$, $-2.3$, 3.2, and 0.6 $M_{\odot}$ (starting from the West beam and counting counterclockwise); and for Beam 5, the corresponding values are $-2.7$, 0.7, 1.8, $-3.6$,  0.8, and $-4.9$ $M_{\odot}$. For Beam 3, the corresponding values are 10.9, 0.4, $-0.9$, 2.3,   5.2, and 4.4 $M_{\odot}$.}These facts indicate that atomic gas flows out from Beam 1, and these atomic mass excesses may be an indicator of accumulated atomic mass loss from the excitation star, $\Delta M_{\ast, \mathrm{a}}$ \citep[e.g.,][]{Lizano+1988}.

The atomic mass excesses of Beam 1 relative to its surrounding beams with a velocity range of $|v_\mathrm{{c,w}} - v| \leq 20$ km s$^{-1}$ are listed in Table \ref{tab:HIflow-properties}, which are set to be two times of the corresponding atomic mass excesses with a velocity range of $v_\mathrm{{c,w}} - v \leq 20$ km s$^{-1}$. The average value of $\Delta M_{\ast, \mathrm{ame}}$ over the five atomic mass excesses (i.e., except for that between Beams 1 and 3, where the atomic mass excesses are significantly different from the other atomic mass excesses) is 6.0 $\pm$ 3.5 $M_{\odot}$ (i.e., $\Delta M_{\ast, \mathrm{a}} \sim$ 6.0 $\pm$ 3.5 $M_{\odot}$), where 3.5 $M_{\odot}$ is the STD of the five atomic mass excesses. The error of atomic mass excess inherited from the original spectrum is far less than 3.5 $M_{\odot}$, and therefore no error of atomic mass excess is presented in Table \ref{tab:HIflow-properties}. $\Delta M_{\ast, \mathrm{a}} \sim \Delta M_{\ast, \mathrm{m}}$, and the outflowing atomic gas would undergo deceleration when it entrained molecular outflow gas, indicating that the velocity associated with $\Delta M_{\ast, \mathrm{a}}$ may be not less than that associated with $\Delta M_{\ast, \mathrm{m}}$ and H~\textsc{i} wind is likely strong enough to drive the molecular outflow.

\subsubsection{Lifetime of the H~\textsc{i} Wind}\label{sec-HI-flow-compare-with-molecular-lifetime}

The lifetime of the H~\textsc{i} wind flow, i.e., the total duration of the flow, $t$, without inclination correction was estimated according to \citet{Lizano+1988}. The distance from the H~\textsc{i} wind to the central excitation star, $l_{\mathrm{HI,wind}}$, was set as the outflow length, $l_{\mathrm{outflow}}$, of the molecular outflow \citep[which, here, corresponds to the red lobe of $^{12}$CO; see][]{Liu+2021}, i.e., $l_{\mathrm{HI,wind}}$ $\sim$ 0.8 pc. The crossing time of the H~\textsc{i} wind, $t_{\mathrm{a}}$, is then $t_{\mathrm{a}} = l_{\mathrm{HI,wind}}/V_{\mathrm{mean,wind}}$ = (1.9 $\pm$ 0.2) $\times 10^4$ yr, where the uncertainty comes only from $V_{\mathrm{mean,wind}}$ because no error of $l_{\mathrm{outflow}}$ was given by \citet{Liu+2021}. Therefore, the stellar mass-loss rate, $\dot{M}_{\mathrm{a}}$, is $\dot{M}_{\mathrm{a}} = M_{\mathrm{HI,wind}}/t_{\mathrm{a}}$ = (12.2 $\pm$ 3.6) $\times$ 10$^{-6}$ $M_{\odot}$ yr$^{-1}$. This value (without inclination correction) is larger than that in HH 7--11 by a factor of $\sim$ 2 and larger than that in L1551 by a factor of $\sim$ 10 \citep[see][]{Lizano+1988, Giovanardi+1992}. The lifetime of the H~\textsc{i} wind is $t = \Delta M_{\ast, \mathrm{a}}/\dot{M}_{\mathrm{a}}$ = (5.3 $\pm$ 3.5) $\times$ 10$^5$ yr, where the accumulated atomic mass, $\Delta M_{\ast, \mathrm{a}}$, is set to 6.0 $\pm$ 3.5 $M_{\odot}$ (see above). All the physical properties above are listed in Table \ref{tab:HIflow-properties}. The dynamical timescale of the red lobe of the molecular outflow, $t_{\mathrm{outflow}}$, as an indicator of the age of the flow, ranges from 1--5 $\times 10^4$ yr traced by four species \citep[i.e., $^{12}$CO, $^{13}$CO, HCO$^+$ and CS; see][]{Liu+2021}. $t$ is about one order of magnitude larger than  $t_{\mathrm{outflow}}$, indicating that the H~\textsc{i} wind is likely strong enough to drive the molecular outflow. Observations with higher spatial resolution of both H~\textsc{i} and molecules are required to further confirm this conclusion.

\begin{figure*}[!htb]
	\centering
	\includegraphics[height=0.45\textwidth]{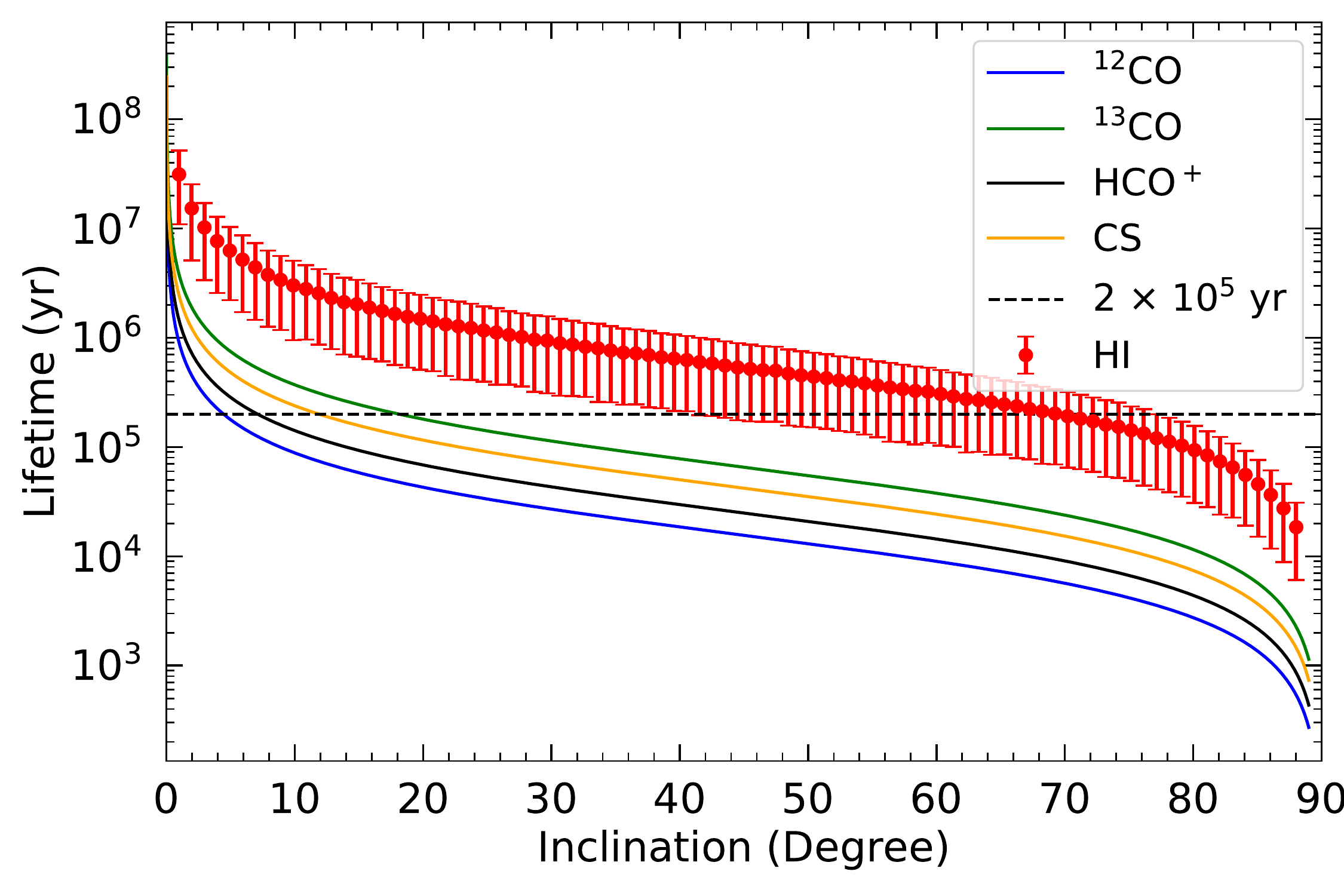}
	\caption{$t$ (red points, where the error bars  indicate the 1$\sigma$ uncertainty) and $t_{\mathrm{outflow}}$ (blue solid line for $^{12}$CO, green for $^{13}$CO, black for HCO$^+$, and yellow for CS) as functions of the inclination. The dashed line is the dynamical timescale of the corresponding star formation \citep[i.e., $\sim$ 2 $\times$ 10$^5$ yr; see][]{Chen+2003}.}
	\label{fig-timescale}
\end{figure*}

To further investigate the influence of the inclination, Figure \ref{fig-timescale} presents both $t$ and $t_{\mathrm{outflow}}$ (including those for the $^{12}$CO, $^{13}$CO, HCO$^+$, and CS outflows) under different inclinations. It can be seen that the lifetime of the H~\textsc{i} wind is much larger than the timescale of the outflow traced by all the four molecules, no matter what inclination is considered. This suggests that the inclination does not change the inequality of $t \gg t_{\mathrm{outflow}}$.

Moreover, \citet{Chen+2003} derived the timescale of the star formation corresponding to this H~\textsc{i} wind or molecular outflow as $\sim$ 2 $\times$ 10$^{5}$ yr by researching the H$_2$ knots surrounding the excitation source. The inclination of this H~\textsc{i} wind or molecular outflow, $\theta_0$, may be $\sim 69^{\circ}\,^{+8^{\circ}}_{-28^{\circ}}$, assuming that the lifetime of the H~\textsc{i} wind equals the timescale of star formation, i.e., $t \sim 2 \times 10^{5}$ yr.

\subsection{Physical Properties with an Inclination of $69^{\circ}$}\label{sec-fix-inclination}

We next calculated the physical properties of the H~\textsc{i} wind assuming that an inclination of $69^{\circ}$ (see Table \ref{tab:HIflow-properties-2}). 
The maximum velocity of the H~\textsc{i} wind changes to $\sim 336.2 \pm 44.6$ km s$^{-1}$. This value is larger than that for HH 7--11 \citep[assuming an inclination of $22^{\circ}$; see][]{Herbig-Jones1983, Lizano+1988} by a factor of $\sim$ 2, and is $\sim$ 60\% of that for L1551 \citep[assuming an inclination of 75$^{\circ}$; see][]{Snell-Schloerb1985, Giovanardi+1992}. The value of the momentum of the H~\textsc{i} wind after inclination correction is 22.66 $M_{\odot}$ km s$^{-1}$, which is larger than that in HH 7--11 by a factor of $\sim 12$ \citep[see][]{Lizano+1988, Giovanardi+1992}, and a factor of $\sim 15$ for that in L1551 \citep[see][]{Giovanardi+1992}. The values of the wind's kinetic energy, i.e., $\sim 2.6 \times 10^{47}$ erg, and the stellar mass-loss rate, i.e., $\sim 3.2 \times 10^{-5}$ $M_{\odot}$ yr$^{-1}$, are larger than those in HH 7--11 by factors of $\sim 28$ and $\sim 10$, and larger than those in L1551 by factors of $\sim 3$ and $\sim 7$, respectively \citep[see these values in HH 7--11 and L1551 in][]{Lizano+1988, Giovanardi+1992}.

The kinetic luminosity of the H~\textsc{i} wind, $L_{\mathrm{HI,wind}} = \displaystyle \frac{1}{2} \dot{M}_{\mathrm{a}} V^2_{\mathrm{max,wind}}$, is 16.0 $\pm$ 10.1 $L_{\odot}$ without inclination correction, which is much less than the bolometric luminosity, $L_{\ast}$, of the excitation source \citep[i.e., 1.7 $\times$ 10$^3$ $L_{\odot}$; see][]{Molinari+2008}. However, the kinetic luminosity of the H~\textsc{i} wind after inclination correction is 322.0 $L_{\odot}$, i.e., $\sim$ 19\% of $L_{\ast}$. The ratio of $L_{\mathrm{HI,wind}}/L_{\ast}$ after inclination correction is comparable to that in HH 7--11 (i.e., 20\%) and in L1551 \citep[i.e., 127\%, see][]{Giovanardi+1992}, while the ratio of $L_{\mathrm{HI,wind}}/L_{\ast}$ before inclination correction is only $\sim$ 1\%. This result implies that an inclination correction is necessary when we calculate the physical properties of the H~\textsc{i} wind.

\begin{deluxetable}{ll}
    \tablenum{4}
	\setlength{\tabcolsep}{35pt}
	\tablecaption{Physical Properties of the H~\textsc{i} Wind for an Inclination of $69^{\circ}$}
	\label{tab:HIflow-properties-2}
	%
	\tablehead{
		\colhead{Quantity}	&	\colhead{Value (1$\sigma$ uncertainty)}}
	\startdata
	$P_{\mathrm{HI,wind}}$ ($M_{\odot}$ km s$^{-1}$) & 22.66 $\pm$ 9.63 \\	
	$V_{\mathrm{max,wind}}$ (km s$^{-1}$) & 336.2 $\pm$ 44.6 \\
	$V_{\mathrm{mean,wind}}$ (km s$^{-1}$) & 112.1 $\pm$ 14.9  \\
	$E_{\mathrm{HI,wind}}$ (erg) & (2.6 $\pm$ 1.2) $\times$ 10$^{47}$  \\			
	$t_{\mathrm{a}}$ (yr) & (7.2 $\pm$ 0.9) $\times$ 10$^3$	\\
	$\dot{M}_{\mathrm{a}}$ ($M_{\odot}$ yr$^{-1}$) & (3.2 $\pm$ 0.9) $\times$ 10$^{-5}$ \\
	$t$ (yr) & (2.0 $\pm$ 1.4) $\times$ 10$^5$\\	
	$L_{\mathrm{HI,wind}}$ ($L_{\odot}$) & 322.0 $\pm$ 202.8
	\enddata
	\tablecomments{See description of each quantity in Table \ref{tab:HIflow-properties}.}
\end{deluxetable}




\section{Summary and Conclusions}\label{sec-summary}

We obtained high-sensitivity H~\textsc{i} spectra with an average rms noise of $\sim$ 7 mK @ 0.1 km s$^{-1}$ toward the high-mass star-forming region G176.51+00.20 using FAST with a 19-beam tracking observational mode. We searched for H~\textsc{i} wind in the central seven beams (i.e., Beams 1--7), but detected it only in Beam 1. The main results are as follows:
\begin{enumerate}
	\item The abundance of H~\textsc{i} in the H~\textsc{i} wind is consistent with that of the HINSA (where the molecular column density is traced by $^{13}$CO). This indicates that there probably exists an internal correlation between the H~\textsc{i} wind and HINSA; such a correlation would enhance the argument of the association between the H~\textsc{i} wind and the molecular outflows. 
	\item The H~\textsc{i} wind is likely strong enough to drive the molecular outflows. This conclusion is not affected by the value of the inclination. 
	\item The mass, momentum, and kinetic energy of the H~\textsc{i} wind and the associated mass-loss rate of the excitation star in G176.51+00.20, a high-mass star-forming region, are about one orders of magnitude larger than those in low-mass star-forming regions (i.e., HH 7--11 and L1551).	
\end{enumerate}

\appendix

\section{Mass and Momentum of the H~\textsc{i} Wind}\label{sec-HIflow-mass-momentum}

A crude ``triangular fit'' was adopted to obtain the relationship between the brightness temperature and the radial velocity \citep[e.g.,][]{Lizano+1988} as
\begin{equation}\label{flow-blue-beam1}
	T_{\mathrm{wind}}(v) = a + b \cdot v,
\end{equation}
where $a$ and $b$ are constants with units of K and K (km s$^{-1}$)$^{-1}$, respectively. The result of the ``triangular fit'' is plotted as the red solid line in Figure \ref{Figure-HIflow-1}(a). Considering that the H~\textsc{i} wind may be contaminated or affected by the ambient environment, and manifesting as an asymmetry between the blue and red lobes, the blue and red lobes were fitted separately with the line center, $v_\mathrm{{c,w}}$, as the velocity of HINSA \citep[see][see also Figure \ref{figure-noise} and the red vertical line in Figure \ref{Figure-HIflow-1}(a)]{Li+2022b}. 

$T_{\mathrm{wind}}$ can be used to calculate the column density of the H~\textsc{i} wind as:
\begin{equation}\label{NHI-flow-red-beam1}
	N_{\mathrm{HI,wind}} = 1.82 \times 10^{18} \int T_{\mathrm{wind}} dv,
\end{equation}
where H~\textsc{i} emission is assumed to be optically thin \citep[][]{Dickey-Benson1982, Lizano+1988, Saha+2018}.

The maximum velocity of the H~\textsc{i} wind, $V_{\mathrm{{max, flow}}}$, reads:
\begin{equation}\label{Vmax-red-beam1}
	V_{\mathrm{{max, flow}}} = - a/b,
\end{equation}
and the mean velocity of the H~\textsc{i} wind is $V_{\mathrm{{mean, flow}}} = V_{\mathrm{{max, flow}}}/3$ for the triangular profile. Considering that the beam size of FAST is $2'.9$ and the distance to the H~\textsc{i} wind is 1.8 kpc, the mass, $M_{\mathrm{HI,wind}}$, and momentum, $P_{\mathrm{HI,wind}}$, of the H~\textsc{i} wind without inclination correction are:
\begin{equation}\label{MHI-flow-red-beam1}
	M_{\mathrm{HI,wind}} = C_0 \int T_{\mathrm{wind}} dv,
\end{equation}
and
\begin{equation}\label{PHI-flow-red-beam1}
	P_{\mathrm{HI,wind}} = C_0 \int |v - v_\mathrm{{c,w}}| T_{\mathrm{wind}} dv,
\end{equation}
respectively, where the constant $C_0$ is:
\begin{equation}\label{MHI-C-constant}
	C_0 = 0.103\; \mathrm{K^{-1}\;} M_{\odot}\; \mathrm{km^{-1} \; s},
\end{equation}
where we assume that the hydrogen mass abundance is 0.74 \citep[see][]{Garden+1991} and the H~\textsc{i} wind emission is from a point source \citep[e.g.,][]{Lizano+1988}. This is reasonable because the value of $V_{\mathrm{{max, flow}}}$ is 120.6 $\pm$ 16.3 km s$^{-1}$ \citep[see][and references therein]{Lizano+1988, Frank+2014}. The value of $C_0$ is larger than that in Equation (\ref{MHI-C-constant}) by a factor of $2\ln 2$ for an extended source with a uniform
brightness temperature over the beam. 
The kinetic energy, $E_{\mathrm{HI,wind}}$, of the H~\textsc{i} wind is \citep[see][]{Giovanardi+1992}:
\begin{equation}\label{EHI-flow-red-beam1}
	E_{\mathrm{HI,wind}} = \frac{1}{2} M_{\mathrm{HI,wind}} V^2_{\mathrm{{max, flow}}}.
\end{equation}

\acknowledgments
This work made use of the data from FAST. FAST is a Chinese national mega-science facility, operated by National Astronomical Observatories, Chinese Academy of Sciences. We would like to thank the anonymous referee for the helpful comments and suggestions that helped to improve the paper. This work was sponsored by the Natural Science Foundation of Jiangsu Province (grant No. BK20210999), the Entrepreneurship and Innovation Program of Jiangsu Province, NSFC grants Nos. 11933011 and 11873019, and the Key Laboratory for Radio Astronomy, Chinese Academy of Sciences.

\facility{FAST, PMO 13.7m}

\software{Astropy \citep{Astropy2013,Astropy2018}, Matplotlib \citep{Matplotlib2007}, Numpy \citep{Numpy2020}, Pandas \citep{Pandas2010, Pandas2021}, Scipy\citep{2020SciPy-NMeth}, Emcee \citep{Foreman-Mackey+2013}}

\end{CJK*}
\end{document}